\documentclass[usenatbib,a4paper]{mn2e}
\usepackage{times}
\usepackage{graphicx}

\title
[CMB constraints on Dark Energy]
{Understanding the origin of CMB constraints on Dark Energy} 

\author[Jassal, Bagla and Padmanabhan]{H. K. Jassal
  $^{1}$, J.~S.~Bagla $^1$
 and   T.~Padmanabhan$^2$  \\
$^{1}$ Harish-Chandra Research Institute, Chhatnag Road,
Jhunsi, Allahabad 211 019, India.\\
$^{2}$ Inter University Centre for Astronomy and Astrophysics,
Post Bag 4, Ganeshkhind, Pune 411 007, India.\\
E-mail: hkj@hri.res.in, jasjeet@hri.res.in, nabhan@iucaa.ernet.in}
 
\begin{document}

\pagerange{\pageref{firstpage}--\pageref{lastpage}} \pubyear{2008}
\maketitle
\label{firstpage}

\begin{abstract}
We study the observational constraints of CMB temperature and polarization
anisotropies on models of dark energy, with special focus on models with
variation in properties of dark energy with time. 
We demonstrate that the key constraint from CMB observations arises from the
location of acoustic peaks. 
An additional constraint arises from the limits on $\Omega_{NR}$ from the
relative amplitudes of acoustic peaks. 
Further, we show that the distance to the last scattering surface is not how
the CMB observations constrain the combination of parameters for models of
dark energy. 
We also use constraints from Supernova observations and show that unlike the
Gold and Silver samples, the SNLS sample prefers a region of parameter space
that has a significant overlap with the region preferred by the CMB
observations.  
This is a verification of a conjecture made by us in an earlier work
\citep{2005PhRvD..72j3503J}.  
We discuss combined constraints from WMAP5 and SNLS observations. 
We find that models with $w \simeq -1$ are preferred for models with
a constant equation of state parameters.
In case of models with a time varying dark energy, we show that
constraints on evolution of dark energy density are almost independent of the
type of variation assumed for the equation of state parameter.  
This makes it easy to get approximate constraints from CMB observations on
arbitrary models of dark energy.
Constraints on models with a time varying dark energy are predominantly due to
CMB observations, with Supernova constraints playing only a marginal role.
\end{abstract}


\section{Introduction}

Observational evidence for accelerated expansion in the universe has been
growing in the last two decades \citep{crisis2, crisis1, crisis3}.  
Independent confirmation using observations of high redshift supernovae
\citep{1998ApJ...509...74G, 1999ApJ...517..565P, nova_data1, nova_data2,
  nova_data3, snls}
has made this result more acceptable to the community.  
Using these observations along with observations of cosmic microwave
background radiation (CMB) \citep{boomerang, wmap_params, wmap5b} and large
scale 
structure \citep{sdss}, we can construct a
``concordance'' model for cosmology and study variations around it (e.g., see
\citet{wmap5a, wmap5b, 2003Sci...299.1532B, 2004ApJ...606..702T}; for an
overview of our current understanding, see \citet{TPabhay, tptalk, tpessay}).

Observations indicate that the dominant component of energy density ---
called dark energy --- should have an equation of state parameter $w\equiv
P/\rho < -1/3$ for the universe to undergo accelerated expansion. 
Indeed, present day observations require $w \simeq -1$.  
The cosmological constant is the simplest explanation for accelerated
expansion
\citep{ccprob,1992ARA&A..30..499C,review3,review1,review2,review4,review5,peri_rev,sami_rev}
and it is known to be consistent with observations.  
In order to avoid theoretical problems related to cosmological constant
\citep{ccprob,1992ARA&A..30..499C}, many other scenarios have been
investigated: these include quintessence
  \citep{quint1,quint2,quint3,quint4,quint5,quint6,quint7,quint8}, k-essence
  \citep{2001PhRvD..63j3510A,k-essence2,k-essence3,k-essence4,k-essence5},
  tachyon  field
  \citep{tachyon5,2002PhRvD..66b1301P,tachyon2,tachyon3,tachyon4,tachyon6,tachyon7,tachyon8,tachyon9,tachyon10,tachyon11,tachyon12}, chaplygin gas  
and its generalisations \citep{gas1,gas2,gas3,gas4}, phantom fields
\citep{phantom1,phantom2,phantom3,phantom4,phantom4a,phantom5,phantom6,phantom7,
phantom9,phantom10,2004PhRvD..70d3539E,2005PhRvD..71f3004N,2007PhLB..646..105B,2007JPhA...40.6835B,2007JETPL..85....1B}, 
branes \citep{brane1,brane2,brane3,brane4,brane5,brane6} and many others 
\citep{water,horz1,horz2,horz3,horz4,horz5,gaussb,viscous,dynamic_new1,dynamic_new2,dynamic_new3,2006PhRvD..74f4021A,2007JHEP...09..048A,arman}    
In these models one can have $w \ne -1$ and in general $w$ varies with
redshift.  
For references to papers that discuss specific models, the reader may consult
one of the many reviews
\citep{review3,review1,review2,review4,review5,alcaniz_rev}.  
Even though these models have been proposed to overcome the fine tuning
problem for cosmological constant, most models require similar fine tuning of
parameter(s) to be consistent with observations. 
Nevertheless, they raise the possibility of $w(z)$ evolving with time (or it
being different from $-1$), which can be tested by observations.

Given that $w$ for dark energy should be smaller than $-1/3$ for
the Universe to undergo accelerated expansion, the energy density of this
component changes at a much slower rate than that of matter and radiation. 
(Indeed, $w=-1$ for cosmological constant and in this case the energy density
is a constant.) 
Unless $w$ is a rapidly varying function of redshift and becomes $w \sim 0$ at
($z \sim 1$), the energy density of dark energy should be negligible at high
redshifts ($z \gg 1$) compared to that of non-relativistic matter.  
If dark energy evolves in a manner such that its energy density is comparable 
to, or greater than the matter density in the universe at high redshifts then
the basic structure of the cosmological model needs to be modified. 
We do not consider such models, instead we confine our attention to
constraints on dark energy in realistic models and choose observations which
are sensitive to evolution of $w(z)$ at redshifts $z \leq 1$.  

Supernova observations permit a large variation in the equation of state
\citep{dynamic_de4,dynamic_de4a}.  
It has recently been argued that a cosmological constant with a small
curvature can be interpreted as a dynamical dark energy model
\citep{2007JCAP...08..011C,virey_etal}.
We have shown that a combination of supernova observations with CMB
observations and  abundance of rich clusters of galaxies provides tight
constraints on variation of dark energy  \citep{jbp,2005PhRvD..72j3503J}. 
Of these CMB data provides the most stringent bounds on the allowed variation
in evolution of dark energy density.
In this paper the main motivation is to study how these models fare in the
light of current CMB data \citep{wmap5a,wmap5b}.

A variety of observations can be used to constrain models of dark energy,
e.g. see \S{II~B} of \cite{2005PhRvD..72j3503J} for an overview. 
Observations of high redshift supernovae provided the first direct evidence
for accelerated expansion of the universe
\citep{1998AJ....116.1009R,1999ApJ...517..565P}. 
This, coupled with the ease with which the high redshift supernova data can be
compared with cosmological models has made it the favourite benchmark for
comparison with models of dark energy.
It is often considered sufficient to compare a model with the supernova data
even though observers and theorists have pointed out potential problems with
the data \citep{astph0303428,astph0506478}
as well as some peculiar implications of the data
\citep{dynamic_de1,dynamic_de2,shashikant}. 
Further, it has been shown that other observations like temperature
anisotropies in the CMB fix the distance to the last scattering surface and
are a reliable probe of dark energy \citep{white,jbp}. 
In this work, we try to understand the origin of the constraint on models of
dark energy from CMB anisotropy observations.
We show that the angular scale of acoustic peaks provides the leading
constraint on the combination of parameters.
An additional constraint comes in through the degeneracy with $\Omega_{NR}$,
where a constraint on $\Omega_{NR}$ from the relative heights of acoustic
peaks translates into an indirect constraint on the equation of state
parameter $w$.

In an earlier work, we compared the constraints on models of dark energy from
supernova and CMB observations and pointed out that models preferred by these
observations lie in distinct parts of the parameter space and there is no
overlap of regions allowed at $68\%$ confidence level
\citep{2005PhRvD..72j3503J}.  
Even though different observational sets are  sensitive to different
combinations of cosmological parameters, we do not expect models favoured by
one observation to be ruled out by another when such a divergence is not
expected. 
This divergence may point to some shortcomings in the model, or to
systematic errors in observations, or even to an incorrect choice of priors.   
We suggested that this may indicate unresolved systematic errors in one of the
observations, with supernova observations being more likely to suffer from
this problem due to the heterogeneous nature of the data sets available
at the time. 
In Supernova Legacy Survey (SNLS) \citep{snls} survey, a concerted effort has
been made to reduce systematic errors by using only high quality observations. 
The systematic uncertainties are reduced by using a single instrument to
observe the fields.  
Using a rolling search technique ensures that sources are not lost and data
is of superior quality (for details see \cite{snls}).
If our claim about Gold+Silver data set were to be true, SNLS data should not
be at variance with the WMAP data\footnote{This has been shown by several
  authors, including ourselves in a much earlier version of this manuscript.} 
In this work we study constraints on dark energy models from high redshift
supernova observations from the SNLS survey and also observations of the
temperature and polarisation anisotropies in the CMB using the WMAP5 data.

This paper is a revised version of an earlier manuscript which became out of
date after WMAP-3 and then WMAP-5 data were released.  In the interim period,
the issue of systematics in inhomogeneous data sets has been accepted,
therefore we do not emphasize that aspect much in this version.  The
focus of the current paper is twofold: to study constraints on dark energy
models in light of the SNLS and WMAP-5 data, and, to understand the
combination of cosmological parameters that is constrained by the CMB
observations. 

\section{Dark Energy}

\subsection{Cosmological equations}

If we assume that each of the constituents of the homogeneous and isotropic
universe can be considered to be an ideal fluid, and  that the
space is flat, the Friedman equations can be written as:
\begin{eqnarray}
\left(\frac{\dot a}{a}\right)^2 &=& \frac{8 \pi G}{3} \rho \\
\frac{\ddot{a}}{a} &=& -\frac{4 \pi G}{3} (\rho + 3P)
\end{eqnarray}
where $P$ is the pressure and $\rho = \rho_{NR}+\rho_{\gamma} + \rho_{_{\rm
 DE}}$ with the respective terms denoting energy densities for nonrelativistic
 matter, for radiation/relativistic matter and for dark energy. 
Pressure is zero for the non-relativistic component, whereas radiation and
relativistic matter have $P_\gamma = \rho_\gamma / 3$.  
If the cosmological constant is the source of acceleration then $\rho_{_{\rm
 DE}} = $~constant and $P_{_{\rm DE}} = - \rho_{_{\rm DE}}$.
Analysis of nonflat cosmologies reveals that allowed range of curvature of the
universe $\Omega_K$  is $-0.012-0.009$ at 95\% confidence level and a  flat
universe is a good fit to the current data \citep{nonflat}. 

An obvious generalisation is to consider models with a constant equation of
state  parameter $w \equiv P/\rho = $~\textit{constant}.    
One can, in fact, further generalise to models with a varying equation of
state parameter $w(z)$.
Since a function is equivalent to an infinite set of numbers (defined e.g. by
a Taylor-Laurent series coefficients), it is clearly not possible to constrain 
the form of an arbitrary function $w(z)$ using a finite number of
observations.  
One possible way of circumventing this issue is to parameterise the function
$w(z)$ by a finite number of parameters and try to constrain these parameters
with the available observational data. 
There have been many attempts to describe varying dark energy with
different parameterisations 
\citep{constraints_2,dynamic_de14,jbp,constraints_10,holographic,Hannestad:2004cb,2005PhRvD..72j3503J}  
where the functional form of $w(z)$ is fixed and the variation is described
with a small number of parameters.   
Observational constraints depend on the specific parameterisation chosen, but
it should be possible to glean some parameterisation independent results from
the analysis.     

To model varying dark energy we use two parameterisations
\begin{equation}
w(z) = w_0 +  w'(z=0) \frac{z}{(1+z)^p} ~ ; ~~~~~~~~ p=1,~2 \label{taylor}
\end{equation}
These are chosen so that, among other things, the high redshift behaviour is
completely different in these two parameterisations
\cite{jbp}.  
If $p=1$, the asymptotic value $w(\infty) = w_0 + w'(z=0)$ and  for
$p=2$, $w(\infty) = w_0$. 
For both $p=1, 2$, the present value $w(0)=w_0$.
Clearly, we must have $w(z \gg 1) \leq -1/3$ for the standard cosmological
models with a hot big bang to be valid. 
This restriction is imposed over and above the priors used in our study. 

The allowed range of parameters $w_0$ and $w'_0\equiv w'(z=0)$ is likely to be
different for different $p$. 
However, the allowed variation at low redshifts in $\rho_{DE}$ should
be similar in both models as observations actually probe the variation of dark
energy density. 

%%%%%%%%%%%%%%%%%%%%%%%%%%%%%%%%%%%%%%%%%%%%%%%%%%%%%%%%%%%%%%%%%%%%%%%%%
\begin{figure}
\begin{center}
\includegraphics[width=2.2in]{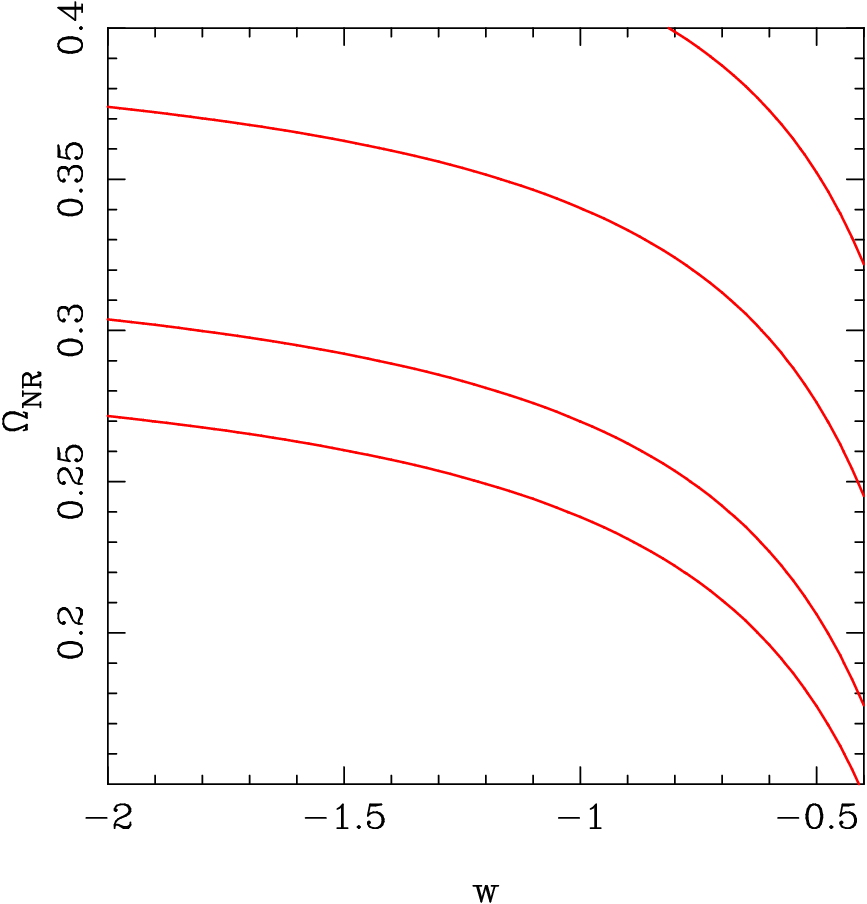}
\end{center}
\caption{This figure shows contours of angular diameter distance. The range
  plotted is the range allowed by WMAP5 data. From the line on to the one at
  bottom  the values correspond to $d_A = 12000,~~13000,~~14273,~~20000$
  and $20100$ Mpc.} 
\end{figure}
%%%%%%%%%%%%%%%%%%%%%%%%%%%%%%%%%%%%%%%%%%%%%%%%%%%%%%%%%%%%%%%%%%%%%%%%%
\begin{figure}
\begin{center}
\includegraphics[width=2.2in]{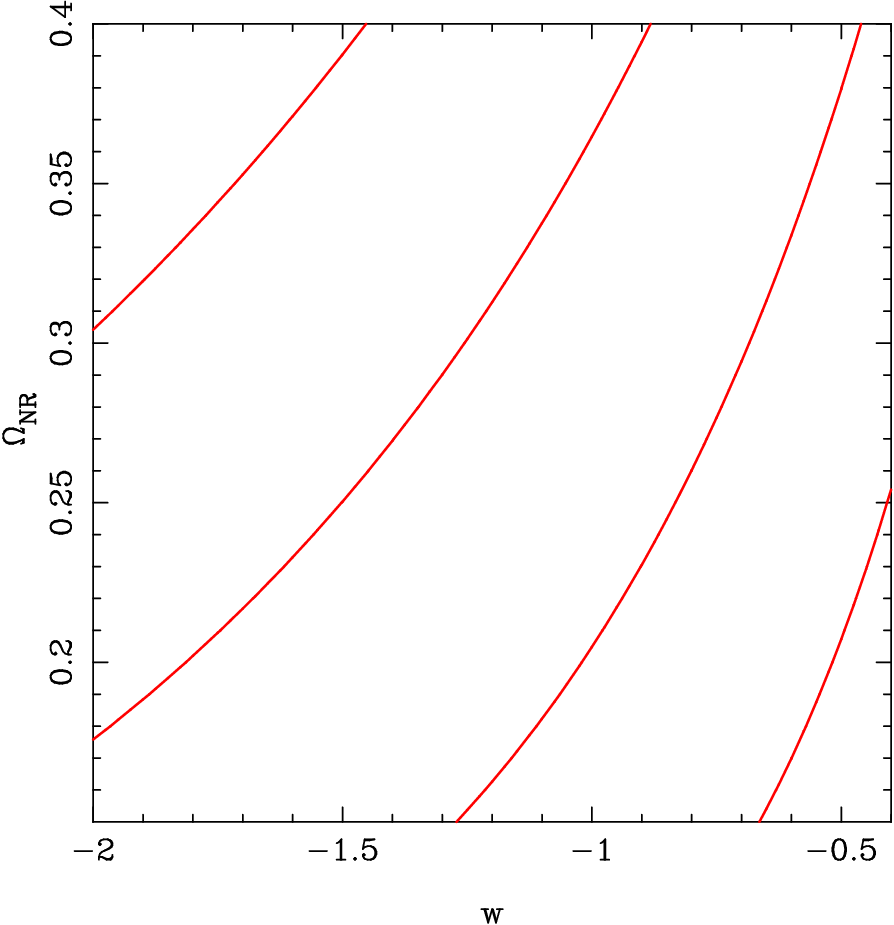}
\end{center}
\caption{This figure shows contours of constant angular size of the Hubble
  radius. The contours from left to right correspond to $\theta^{-1} = 0.0165,~~0.017,~~0.18,~~0.020$.}
\end{figure}
%%%%%%%%%%%%%%%%%%%%%%%%%%%%%%%%%%%%%%%%%%%%%%%%%%%%%%%%%%%%%%%%%%%%%%%%%
%%%%%%%%%%%%%%%%%%%%%%%%%%%%%%%%%%%%%%%%%%%%%%%%%%%%%%%%%%%%%%%%%%%%%%%%%%
\begin{figure*}
\begin{center}
\begin{tabular}{ccc}
\includegraphics[width=2.2in]{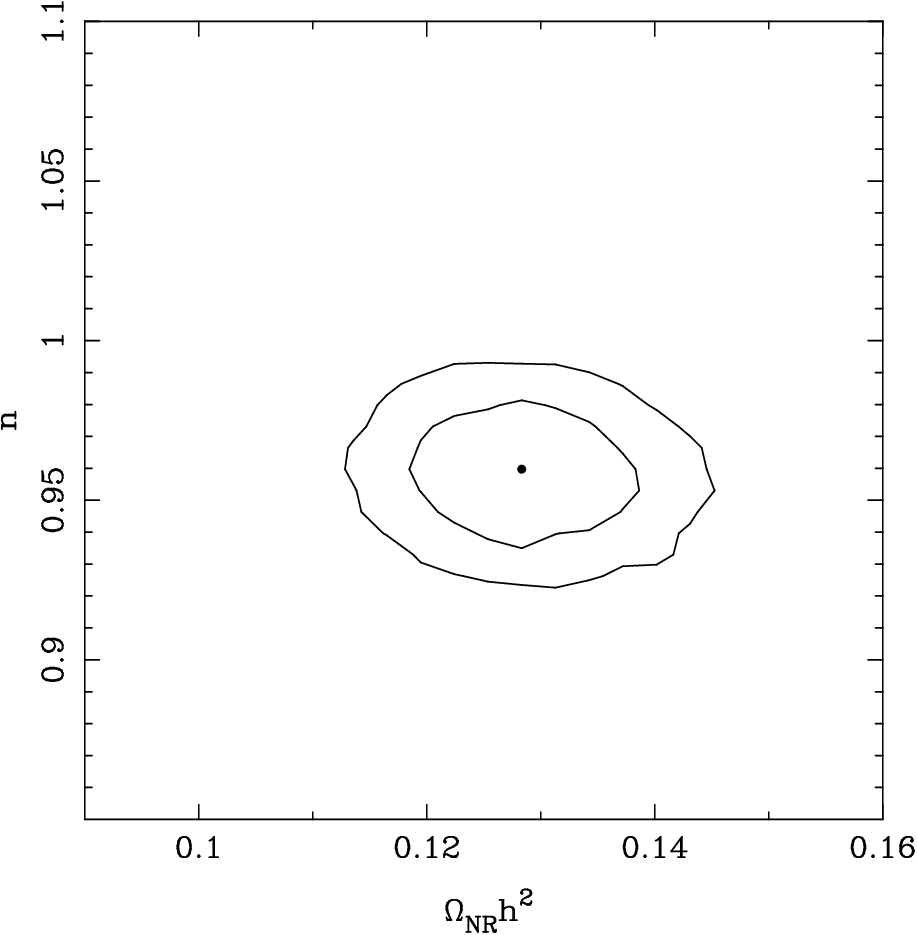} &
\includegraphics[width=2.2in]{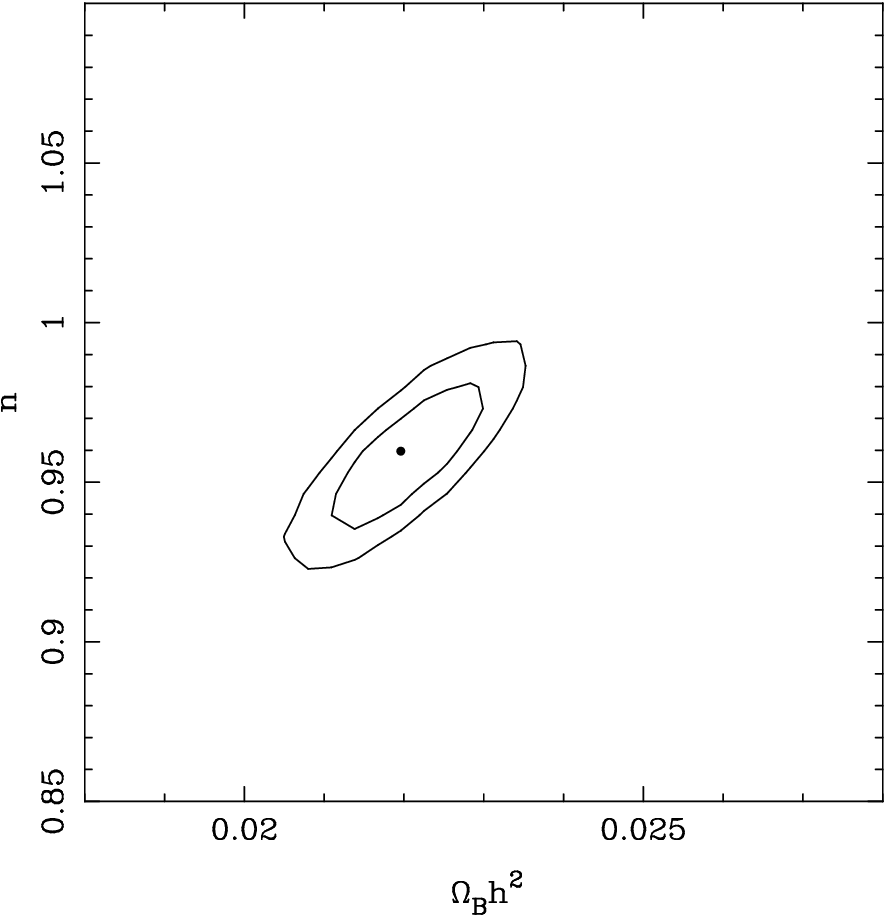} &
\includegraphics[width=2.2in]{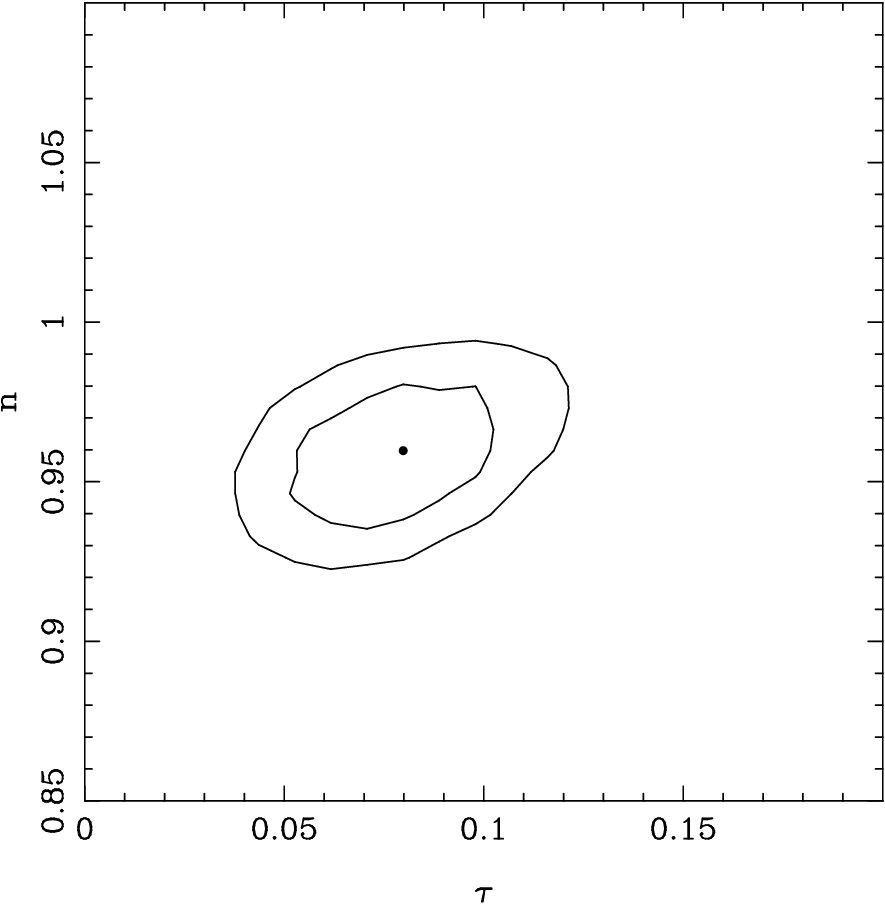} 
\end{tabular}
\end{center}
\caption{Marginalized likelihood contours for different parameters for
  $\Lambda$CDM model.  The regions enclosed by the contours are 68\% and 95\%
  confidence limits. The results are consistent with WMAP5 results for
  $\Lambda$CDM model. The left plot shows the allowed range in
  $n-\Omega_{NR}h^2$ plane. The next figure shows the correlation between
  parameters $n$ and $\Omega_B h^2$. The figure on the right shows
corresponding contours in $n-\tau$ plane.}
\end{figure*}
%%%%%%%%%%%%%%%%%%%%%%%%%%%%%%%%%%%%%%%%%%%%%%%%%%%%%%%%%%%%%%%%%%%%%%%%%%

\subsection{Observational Constraints}

\subsubsection{Supernova Data}

In this work, we concentrate on SN and WMAP observations. 
SN data provides geometric constraints for dark energy evolution.
These constraints are obtained by comparing the predicted luminosity
distance to the SN with the observed one.
The theoretical model and observations are compared for luminosity
measured in magnitudes:
\begin{equation}
m_{B}(z)={\mathcal M} + 5 log_{10}(D_{L})
\end{equation}
where ${\mathcal M}=M-5log_{10}(H_0)$ and $D_{L}=H_{0}d_{L}$, $M$
being the absolute magnitude of the object and $d_L$ is the luminosity
distance  
\begin{equation}
d_{L}=(1+z) a(t_0)r(z);~~~~r(z)=c \int \frac{dz}{H(z)}  
\end{equation}
where $z$ is the redshift.
This depends on evolution of dark energy through $H(z)$. 
For our analysis we use the SNLS data set \citep{snls} and for reference we
also used the combined {\it gold} and {\it silver} SN data set in
\cite{nova_data3} (see also \citep{peri_snls}). 
This data is a  collection of supernova observations from
\cite{nova_data1,1998ApJ...509...74G} and many other sources with $16$
supernovae discovered with Hubble space telescope \citep{nova_data3}. 
The parameter space for comparison of models with SN observations is
small and we do a dense sampling of the parameter space.

\subsubsection{CMB Data}

CMB anisotropies constrain dark energy in two ways, through the distance
to the last scattering surface and through the Integrated Sachs Wolfs (ISW)
effect \citep{isw0}.   
Given that the physics of recombination and evolution of perturbations
does not change if $w(z)$ remains within some {\it safe limits}, any
change in the location of peaks will be due to dark energy \citep{white}.   
For models with a variable $w(z)$, the constraint is essentially on an
effective value $w_{eff}$ \citep{2005PhRvD..72j3503J}. 
This constraint can arise either through the angular diameter distance or the
angular size of the acoustic horizon seen reflected in the scale corresponding
to the acoustic peaks in the CMB angular power spectrum. 

Figure~1 shows contours of equal angular diameter distance to the last
scattering surface in the $\Omega_{nr}-w$ plane. 
We assumed a fixed value of $H_0$ and $\Omega_B$ for this plot.
If the distance to the last scattering surface is the key constraint on models
of dark energy then the likelihood contours should run along the contours of
constant distance in this plane.

We may use the angular size of the Hubble radius at the time of decoupling as
an approximate proxy for the angular size of the acoustic horizon for the
purpose of this discussion.
The approximate angular size $\theta$ of the Hubble radius at the time of
decoupling can be written as:
\begin{eqnarray}
\theta^{-1} &=& \frac{H_0/H(z)}{\int\limits_0^z
  {dy}/{\left(H(y)/H_0\right)}}
\nonumber \\
 &\simeq & \frac{\left({\Omega_{NR}
     \left(1+z\right)^3 }\right)^{-1/2}}{\int\limits_0^z
   {dy}/{\sqrt{ \Omega_{NR} \left(1+z\right)^3 +
     \varrho^{DE}(z)/\varrho^{DE}_0 }}}  \nonumber \\
 & \equiv &  \frac{\left({\Omega_{NR}
     \left(1+z\right)^3 }\right)^{-1/2}}{\int\limits_0^z
   {dy}/{\sqrt{ \Omega_{NR} \left(1+z\right)^3 +
  \Omega_{de} \left(1+z\right)^{3 \left( 1 + w_{eff} \right)}}}}  .
\label{eqn:weff}
\end{eqnarray}
Clearly, the value of the integral will be different if we change $w_0$,
$w'(z=0)$ and there will also be some dependence on the parameterized form.  
If the location of peaks in the angular power spectrum of the CMB
provide the main constraint, this can only constrain $w_{eff}$ and not all of
$w_0$, $w'(z=0)$ and $p$.  
Therefore if the present value $w_0 < w_{eff}$ then it is essential that
$w'(z=0) > 0$, and similarly if $w_0 > w_{eff}$ then $w'(z=0) < 0$ is needed
to ensure that the integrals match. 
Specifically, the combination of $w_0$, $w'(z=0)$ and $p$ should give us 
$w_{eff}$ within the allowed range.  

Figure~2 shows contours of constant $\theta$ in the $\Omega_{NR}-w$ plane.  
If the angular scale of acoustic peaks is the key constraint arising from CMB
observations for models of dark energy then we should see two features in the
likelihood contours:
\begin{itemize}
\item
Likelihood contours for models with a constant equation of state parameter $w$
in the $\Omega_{NR}-w$ plane should run along contours of constant $\theta$,
as shown in Figure~2.
\item
Likelihood contours for different models of varying dark energy should
coincide in the $\Omega_{NR}-w_{eff}$ plane.
These should also coincide with the contours for models with a constant $w$ in
the $\Omega_{NR}-w$ plane.
\end{itemize}
The origin of the CMB constraints on dark energy therefore is not in the raw
distance to the surface of last scattering but in the combination of
parameters that determines the location of peaks in the angular power
spectrum.  
It is important to note that the distance to the surface of last scattering is
a derived quantity. 
The CMB observations constrain only one number, the effective equation of
state.  
There is no ambiguity except for models with early dark energy
\citep{linder_0803}.
In these models, the growth of perturbations is slower than models in which
dark energy comes into play at late times \citep{bena_ber,2006JCAP...06..026D}.

%%%%%%%%%%%%%%%%%%%%%%%%%%%%%%%%%%%%%%%%%%%%%%%%%%%%%%%%%%%%%%%%%%%%%%%
\begin{figure}
\begin{center}
\includegraphics[width=2.2in]{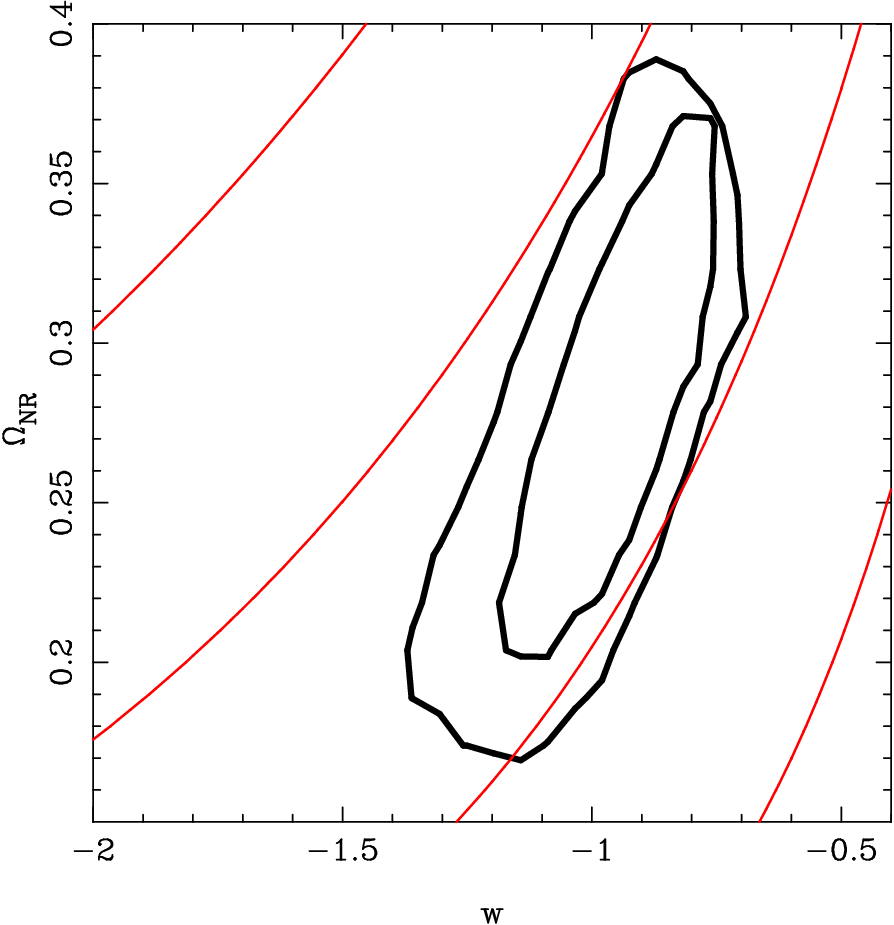}
\end{center}
\caption{This figure shows  contours of angular diameter distance
  (blue/dashed) and of constant angular size of the Hubble radius (red/solid)
  overlaid with likelihood regions allowed by WMAP5 data in the
  $\Omega_{NR}-w$ plane.} 
\end{figure}
%%%%%%%%%%%%%%%%%%%%%%%%%%%%%%%%%%%%%%%%%%%%%%%%%%%%%%%%%%%%%%%%%%%%%%%

In our analysis, we use the angular power spectrum of the CMB temperature
anisotropies \citep{cmbrev1,cmbrev2,kandu} as observed by WMAP
\citep{wmap5a,wmap5b} 
and these are compared to theoretical predictions using the likelihood program
provided by the WMAP team \citep{wmap5a,wmap5b}. 
We vary the amplitude of the spectrum till we get the best fit with WMAP
observations.
The CMBFAST\footnote{http://www.cmbfast.org} package
\citep{1996ApJ...469..437S} is used for computing the 
theoretical angular power spectrum for a given set of cosmological
parameters.
We have combined the likelihood program with the CMBFAST code and this
required a few minor changes in the CMBFAST driver routine.
We also made changes in the driver program to implement Monte Carlo Markov
Chain for sampling the parameter space.  
Please see \citep{2005PhRvD..72j3503J} for details of the MCMC implementation. 

Although we can use other observations like abundance of rich clusters,
baryonic features in the power spectrum, etc. but we find that the two
observations used here are sufficient for this study \citep{jassal}.

\section{Results}

In this section we will describe the results of our study.  
We studied models in three classes:  
\begin{itemize}
\item
Models with a constant equation of state parameter $w$.  
We studied models with perturbations in dark energy
\citep{perturb1, perturb2, perturb3, perturb4, 2007JCAP...02..016F,
  2006PhRvD..74b7301F, perturb5, hkj_depert,pert1,2004A&A...421...71M,
  2004PhRvD..70h3003G,2005PhRvD..71l3505G,
  2006MNRAS.368..751N,2008arXiv0809.3349S, 2005PhRvD..71d7301H,pert2,pert3} as well
as without.     
\item
Models with a varying equation of state parameter $w$, with variation given by
Eqn.3 (p=1).  Perturbations in dark energy were not taken into account in
this case.
\item
Models with a varying equation of state parameter $w$, with variation given by
Eqn.3 (p=2).  Perturbations in dark energy were not taken into account in
this case too.
\end{itemize}
We analyse the allowed range of cosmological parameters for these cosmologies
and consider the probability with which the $\Lambda$CDM model is allowed
within these three classes of models. 
In light of the significant disagreement between the allowed range of
parameters from the high redshift supernova data from the Gold+Silver set
and the CMB anisotropies from WMAP observations
\citep{2005PhRvD..72j3503J,peri_snls}, we also check the degree of overlap
between the parameter space allowed by the supernova and the CMB observations
respectively.  
The newly released 'Union' dataset includes data from the Supernova
Legacy Survey, the ESSENCE Survey \citep{2007ApJ...666..694W} and
the extended distant supernova dataset from HST along with the older datasets
\citep{uniondata}. 
The combined data favours variation in dark energy equation of state.

%%%%%%%%%%%%%%%%%%%%%%%%%%%%%%%%%%%%%%%%%%%%%%%%%%%%%%%%%%%%%%%%%%%%%%%
\begin{table}
\label{tab:priors}
\caption{This table lists the priors used in the present work.  
Apart from the range of parameters listed in the table, we assumed
that the universe is flat.
We assumed that the primordial power spectrum had a constant index. 
Further, we ignored the effect of tensor perturbations. 
The range of values for $w_0$ and $w'(z=0)$ is as given below, but
with the constraint that $w(z=1000) \leq -1/3$.  
Any combination of $w_0$ and $w'(z=0)$ that did not satisfy this
constraint was not considered. 
Values given in parenthesis were used for analysing constraints from high
redshift supernovae in \S{3.1}.}
\begin{center}
\begin{tabular}{||l|l|l||}
\hline
\hline
Parameter & Lower limit & Upper limit  \\
\hline
\hline
$\Omega_B$ & $0.03$ & $0.06$  \\
\hline
$\Omega_{NR}$ & $0.1$ & $0.5 (0.7)$  \\
\hline
$h$ & $0.6$ & $0.8$  \\
\hline
$\tau$ & $0.0$ & $0.4$  \\
\hline
$n$ & $0.86$ & $1.10$ \\
\hline
$w_0$ & $-2.0 (-4.0)$ & $-0.4$  \\
\hline
$w'(z=0)$ & $-5.0$ & $5.0$ \\
\hline
\hline
\end{tabular}
\end{center}
\end{table}
%%%%%%%%%%%%%%%%%%%%%%%%%%%%%%%%%%%%%%%%%%%%%%%%%%%%%%%%%%%%%%%%%%%%%%%
%%%%%%%%%%%%%%%%%%%%%%%%%%%%%%%%%%%%%%%%%%%%%%%%%%%%%%%%%%%%%%%%%%%%%%%
\begin{figure*}
\begin{center}
\begin{tabular}{cc}
\includegraphics[width=2.2in]{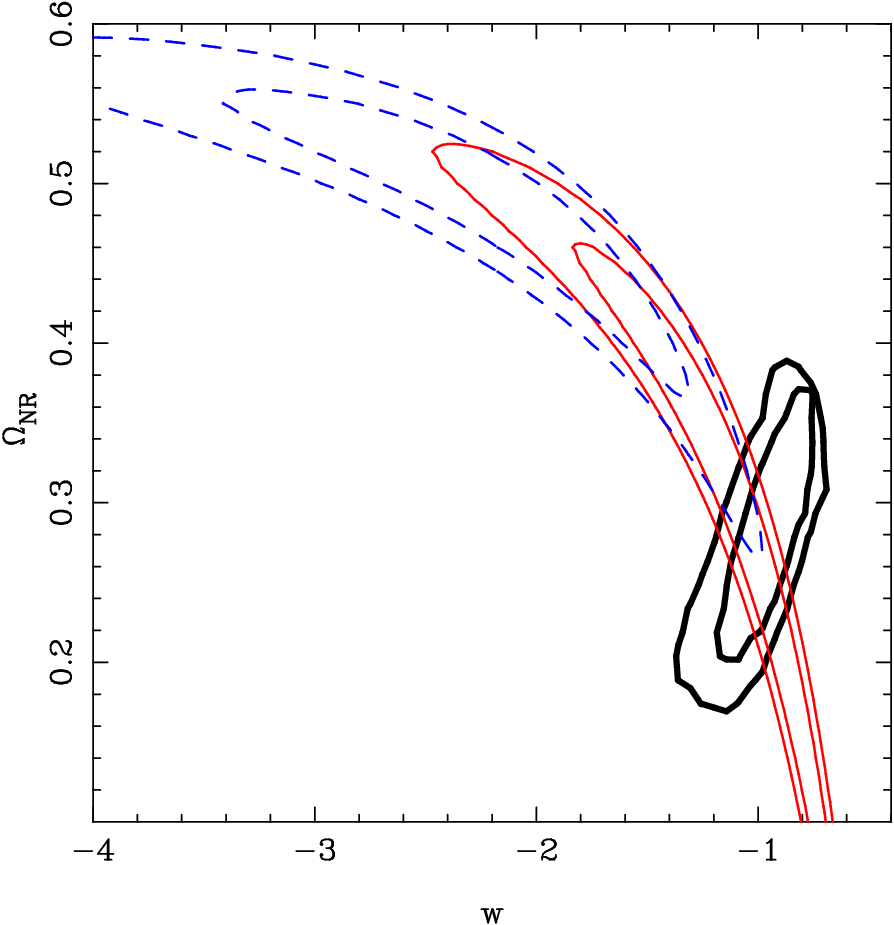} 
&
\includegraphics[width=2.2in]{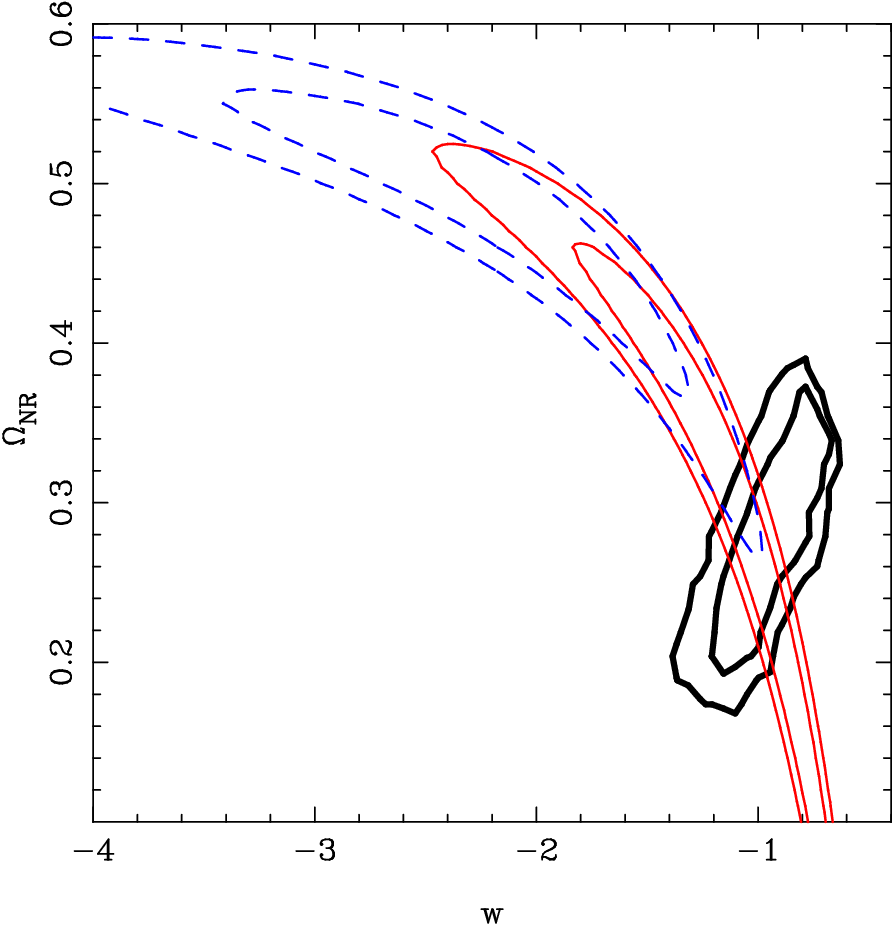} 
\\
\includegraphics[width=2.2in]{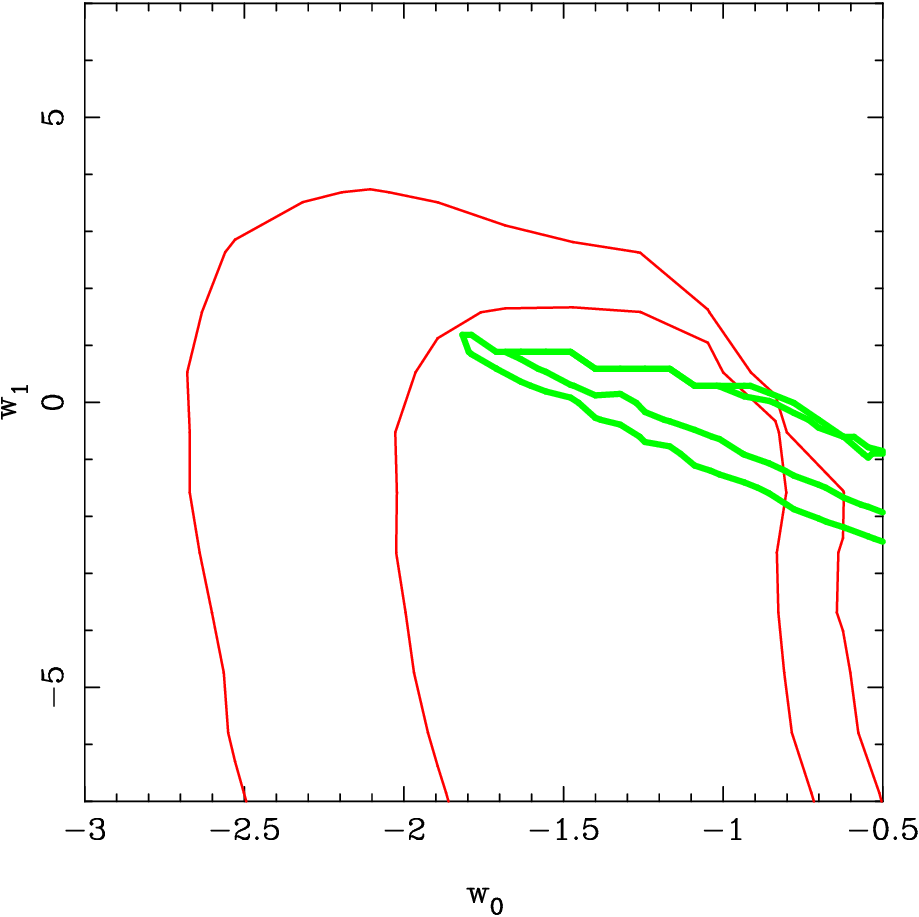} 
&
\includegraphics[width=2.2in]{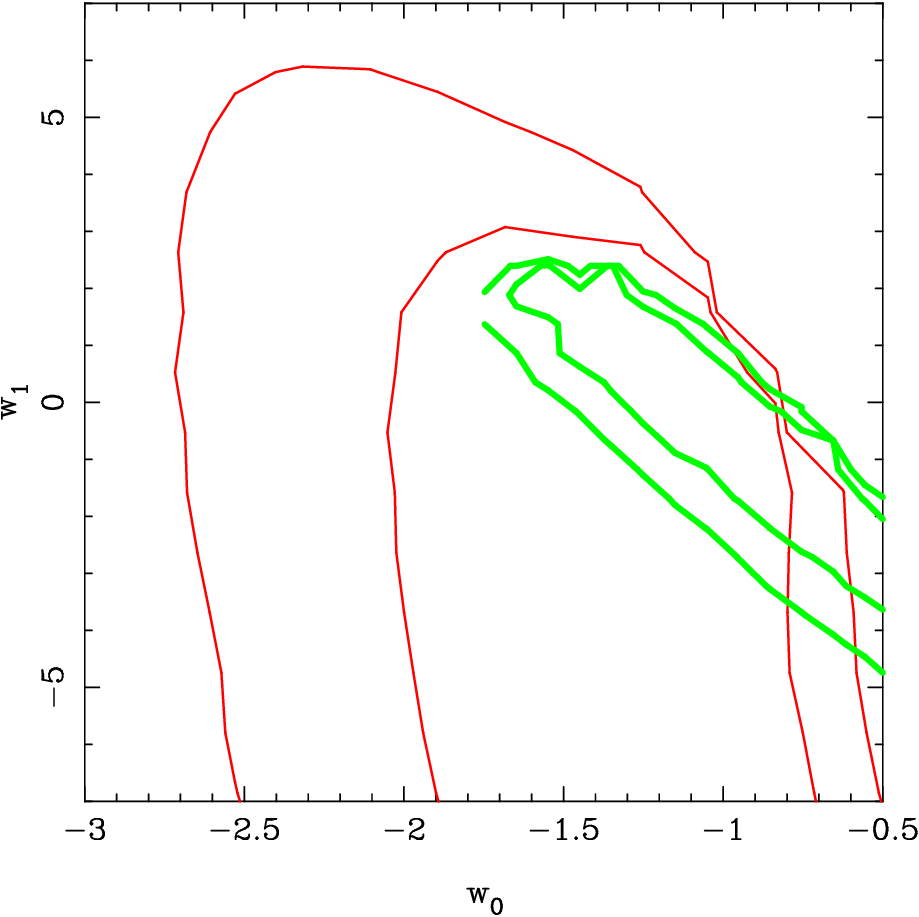}
\end{tabular}
\end{center}
\caption{Marginalized likelihood contours in $\Omega_{NR}-w$ plane for
  different models. In the top figures the blue dashed lines correspond to
  68\%, 95\% confidence level using the gold data. The red solid lines
  correspond to confidence levels in SNLS data. The black thick lines are
  marginalized confidence levels using WMAP5 data. The top left panel if for
  constant $w$ models with a homogeneous dark energy and the one on the right
  is when we include dark energy perturbations. The bottom left plot is for
  models with $p=1$ and the right plot is with $p=2$.}
\end{figure*}
%%%%%%%%%%%%%%%%%%%%%%%%%%%%%%%%%%%%%%%%%%%%%%%%%%%%%%%%%%%%%%%%%%%%%%%

\subsection{Cosmological Constant}

We begin with a very brief review of constraints on parameters in the case
where a cosmological constant is the source of accelerating expansion of the
universe. 
We used priors given in Table~1, except that the equation of state parameter
is fixed to $w=-1$.
Constraints on cosmological parameters are listed in Table~2.  
We would like to note that our results match those obtained by other authors
\citep{wmap5a,wmap5b}.
Figure~3 shows contours of likelihood for some pairs of parameters as an
example.  
We have shown contours in the $n-\Omega_{NR}h^2$, $n-\Omega_B h^2$, and,
$n-\tau$ plane.
we see that there is a strong correlation between $n-\Omega_B h^2$.
These likelihood contours have also been shown as a reference for equivalent
plots for models with $w \neq -1$, and help in checking the effect of the
additional dark energy parameters on allowed range of other parameters.

\subsection{Constant w}

We first evaluate the nature of the CMB constraint on models of dark energy.  
Figure~4 shows likelihood contours from WMAP5 observations in the
$\Omega_{NR}-w$ plane for constant $w$ models.  
We find that the orientation of these contours is roughly along contours of
constant $\theta$. 
To illustrate this, we have overlaid contours from Figure~1 and Figure~2 in
Figure~4.  
On the other hand there is no similarity between the likelihood and contours
of distance to the last scattering surface. 
Thus we may conclude that the dominant constraint provided by the CMB
observations arises from the location of peaks in the angular power spectrum. 
The reason for this is that the angular diameter distance to the last
scattering surface is a derived quantity, whereas the location of
peaks in the angular power spectrum of temperature anisotropies is a direct
observable. 

We use priors given in Table~1 for models with a constant equation of state
parameter, with the obvious constraint that $w'(z=0)=0$. 
For supernova observations, we used wider priors for $w$ and $\Omega_{NR}$ in
order to illustrate the differences between the two data sets studied here.  
We begin with a brief summary of results for the Gold+Silver data set.  
The best fit model in this case is $w=-1.99$ and $\Omega_{NR}=0.47$. 
The allowed range for $w$ at $95\%$ confidence limit for large priors is
$-3.73 \leq w \leq -1.25$. 
The corresponding range for the density parameter is $0.28 \leq \Omega_{NR}
\leq 0.57$. 
With SNLS data, the best fit model is $w=-1.06$ and $\Omega_{NR} = 0.29$.  
The allowed range for $w$ at $95\%$ confidence limit is $-2.36 \leq w \leq -
0.74$.  
The allowed range for the density parameter is $0.11 \leq \Omega_{NR}
\leq 0.48$.  
There is clearly a large shift in the allowed values of parameters. 

We illustrate this in Figure~5 (top-left panel) where we have plotted the
regions allowed by the two data sets at $68\%$ confidence levels in the
$w-\Omega_{NR}$ plane.  
Dashed line shows the region allowed by the Gold+Silver data set and the solid
line is for the SNLS data set.  
We can deduce the following from this figure:
\begin{itemize}
\item
The region allowed by these two data sets at $68\%$ confidence level has some
overlap, thus we may say that the two sets are consistent with each
other.  
\item
The overlap is at $\Omega_{NR} \geq 0.36$ and is thus at margins of what is
allowed by other observations. 
\item
The $\Lambda$CDM model is ruled out at $68\%$ confidence level by the
Gold+Silver data set. 
\item
The best fit of each set is ruled out by the other data set at this confidence
level.  
Indeed, the best fit of Gold+Silver data set is allowed by the SNLS data with
a probability $\mathcal{P}=12.65\%$ while the best fit of the SNLS data set is
allowed by the Gold+Silver data set with  $\mathcal{P}=8.14\%$. 
\end{itemize}
This point is reiterated by the likelihoods of $w$ and $\Omega_{NR}$ for these
models in the same panel.

The figure shows the large overlap between the likelihood curves corresponding
to SNLS  data and WMAP data where the Gold+Silver data clearly favours higher
values of $\Omega_{NR}$ and more negative $w$. 
The phantom models are still allowed but the SNLS data as well as WMAP data
show a preference for models close to a cosmological constant.

For comparison, WMAP allows $-1.25 \leq w \leq -0.7$ and $0.2 \leq
\Omega_{NR} \leq 0.38$ if dark energy is assumed to be smooth. 
If we allow for perturbations in dark energy then the limits on the equation
of state parameters changes to $-1.25 \leq w \leq -0.64$ and $0.20 \leq
\Omega_{NR} \leq 0.38$ as shown in the top right panel of Figure~5.

These figures allow us to conclude that:
\begin{itemize}
\item
WMAP observations of temperature and polarization anisotropies strongly
favour models around $w=-1$, i.e., the $\Lambda$CDM model. 
As a result, WMAP and Gold+Silver data sets have a small region of overlap as
the latter does not favour models around $w=-1$.
(It is this disagreement that had led us to suggest that the supernova data
set could be plagued by some systematic effects \citep{2005PhRvD..72j3503J}, 
particularly as it contains supernovae from a number of different sources.  In
that work, we have used WMAP first year data.) 
\item
WMAP and SNLS data sets have a region of overlap within $68\%$ confidence
levels.   
\item
There is no significant change in the likelihood contours for other
cosmological parameters as we go from the cosmological constant model to dark
energy with a constant equation of state parameter (not constrained to
$w=-1$), or when we go from a smooth dark energy to the model where dark
energy is allowed to cluster.
\end{itemize}
Thus we can say that SNLS and WMAP data are in (much) better agreement as
compared to the Gold+Silver and WMAP data sets. 

\subsection{Varying $w(z)$}

It has been claimed that observations, in particular observations of high
redshift supernovae (the Gold and Gold+Silver data sets) favour evolution of
dark energy \citep{dynamic_de4,dynamic_de4a,dynamic_de1,dynamic_de6,dynamic_de7,dynamic_de8,dynamic_de9,dynamic_de13,dynamic_de14,dynamic_de15,dynamic_de11,dynamic_de12,dynamic_de2,dynamic_de10,dynamic_de5}. 
As such a variation is impossible if acceleration of the universe is caused by
the cosmological constant, it is important to test this claim.  
Note that the term ``Evolution of dark energy'' has been used for evolution of
the equation of state parameter, as well as for evolution of energy density
for the dark energy component.
In an earlier study using the Gold+Silver data set, we had found that
supernova observations {\it do not} favour evolution of the equation of state
parameter over models with a constant $w \ll -1$. 
But these models are favoured strongly as compared to the cosmological
constant model, which was allowed with $\mathcal{P}=6.3\%$ amongst models with
constant $w$.   
When combined with WMAP and other constraints, the allowed variation of dark
energy is restricted to a narrow range and models around the Cosmological
constant are favoured \citep{2005PhRvD..71j3515S,2005PhRvD..72j3503J}. 
We should note that if we combine only the Gold+Silver (or Gold) supernova and
WMAP data then results favour evolution of $\rho_{DE}$, but adding
observations of galaxy clustering removes this inclination. 

%%%%%%%%%%%%%%%%%%%%%%%%%%%%%%%%%%%%%%%%%%%%%%%%%%%%%%%%%%%%%%%%%%%%%%%%%

\begin{figure*}
\begin{center}
\begin{tabular}{cc}
\includegraphics[width=2.2in]{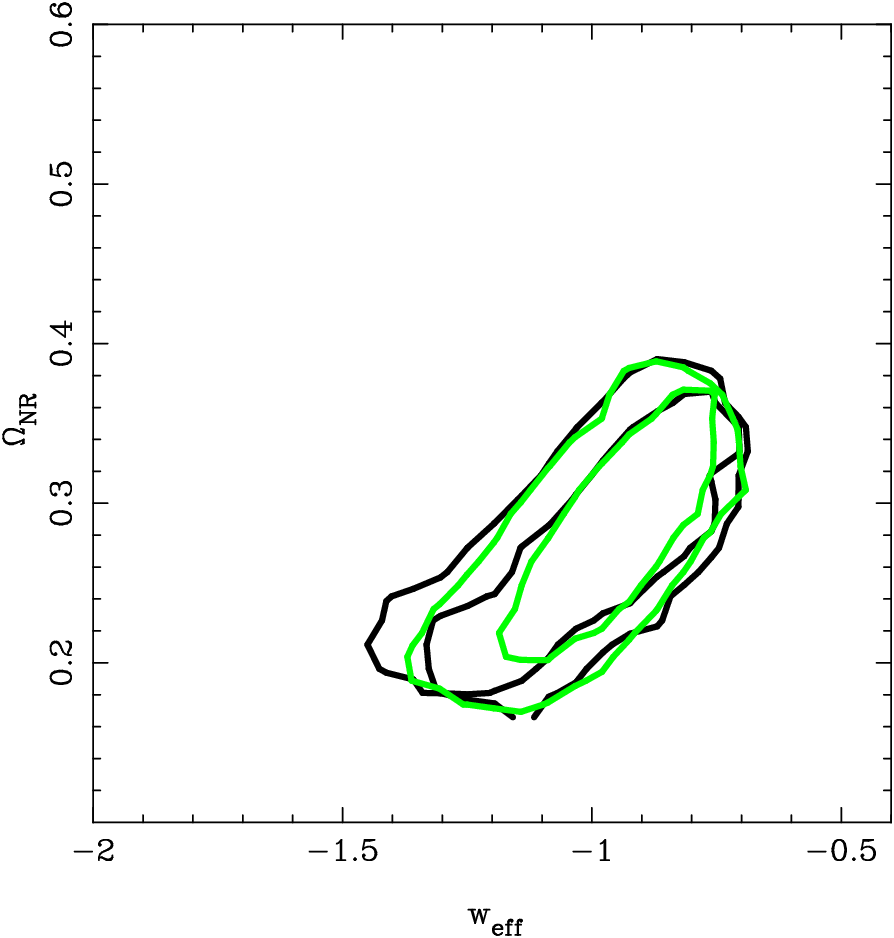}
&
\includegraphics[width=2.2in]{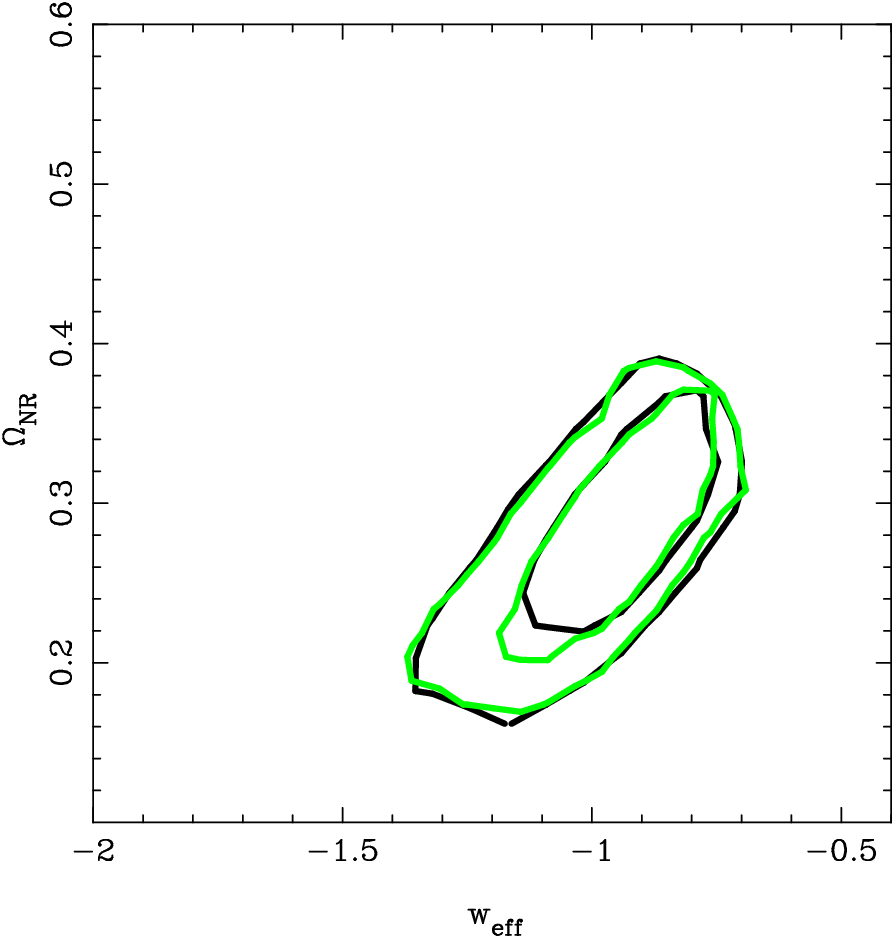}
\end{tabular}
\end{center}
\caption{The black contours in this figure show constraints on the effective
  dark energy equation of state   $w_{eff}$ for   $p=1$ and $p=2$  from the
  WMAP5 data.   The green lines are constraints on    constant $w-\Omega_{NR}$
  (without perturbations). These clearly show that the 
  CMB data constrains effective equation of state at the last scattering
  surface.}
\end{figure*}
%%%%%%%%%%%%%%%%%%%%%%%%%%%%%%%%%%%%%%%%%%%%%%%%%%%%%%%%%%%%%%%%%%%%%%%
\begin{figure*}
\begin{center}
\begin{tabular}{ccc}
\includegraphics[width=1.9in]{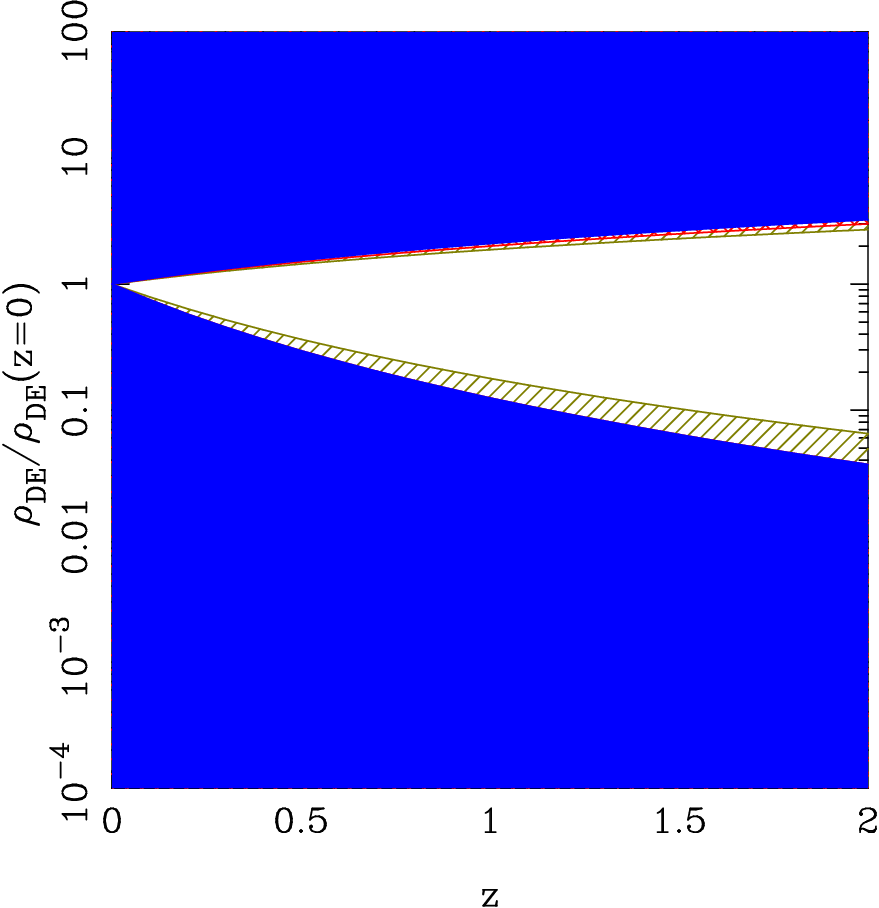} 
&
\includegraphics[width=1.9in]{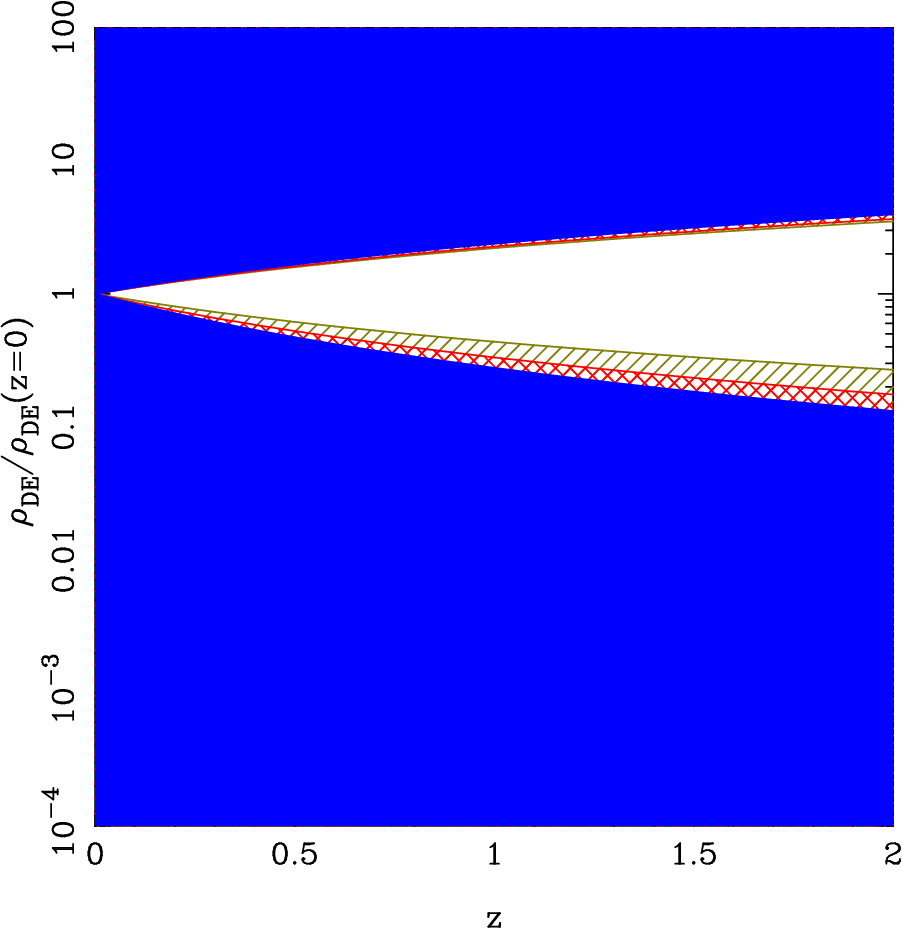} 
&
\includegraphics[width=1.9in]{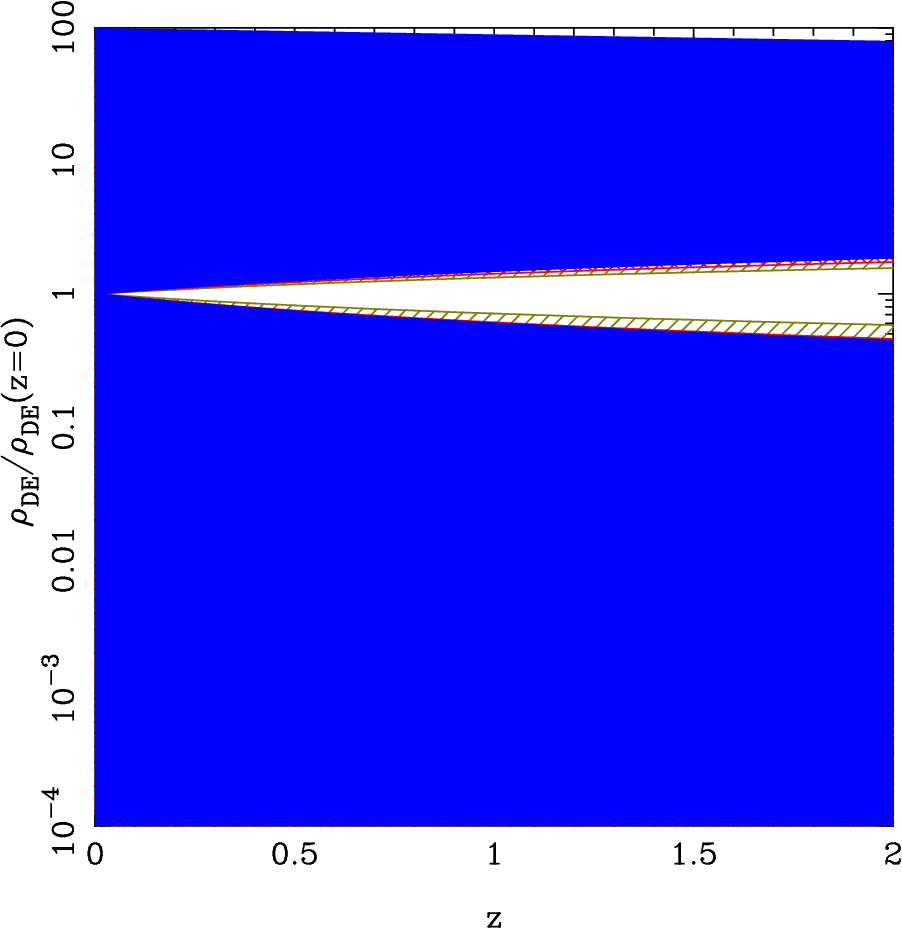} 
\\
\includegraphics[width=1.9in]{fig10a.ps} 
&
\includegraphics[width=1.9in]{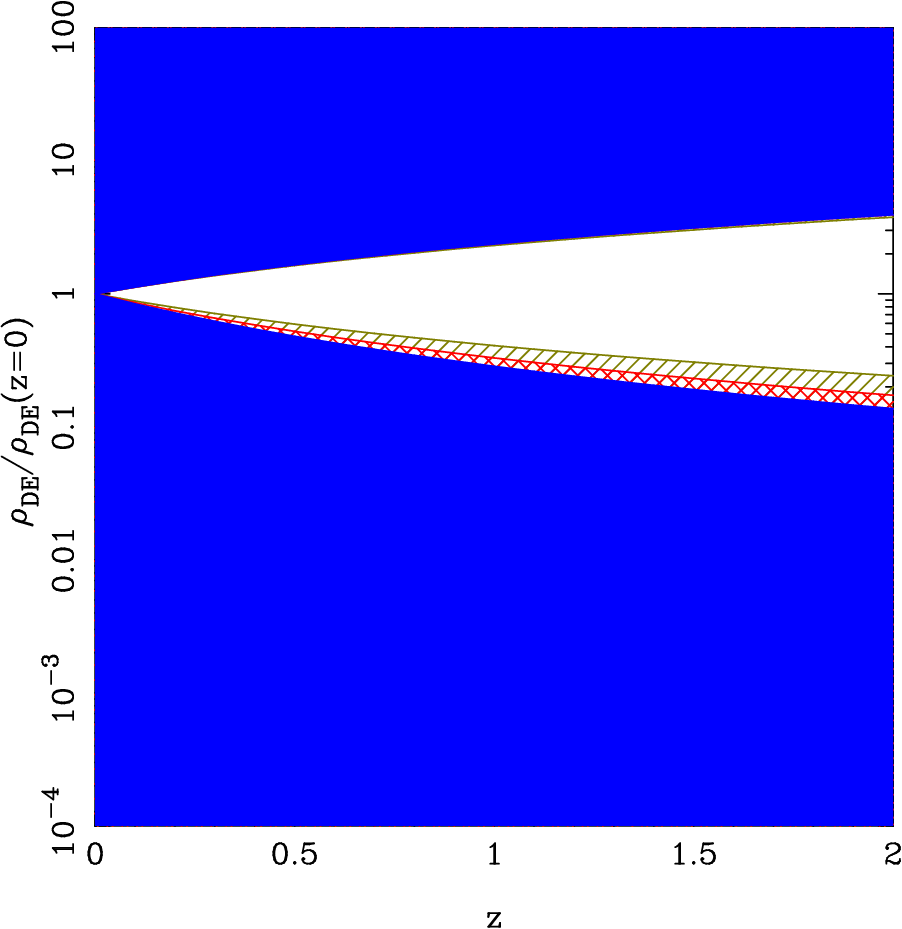}
&
\includegraphics[width=1.9in]{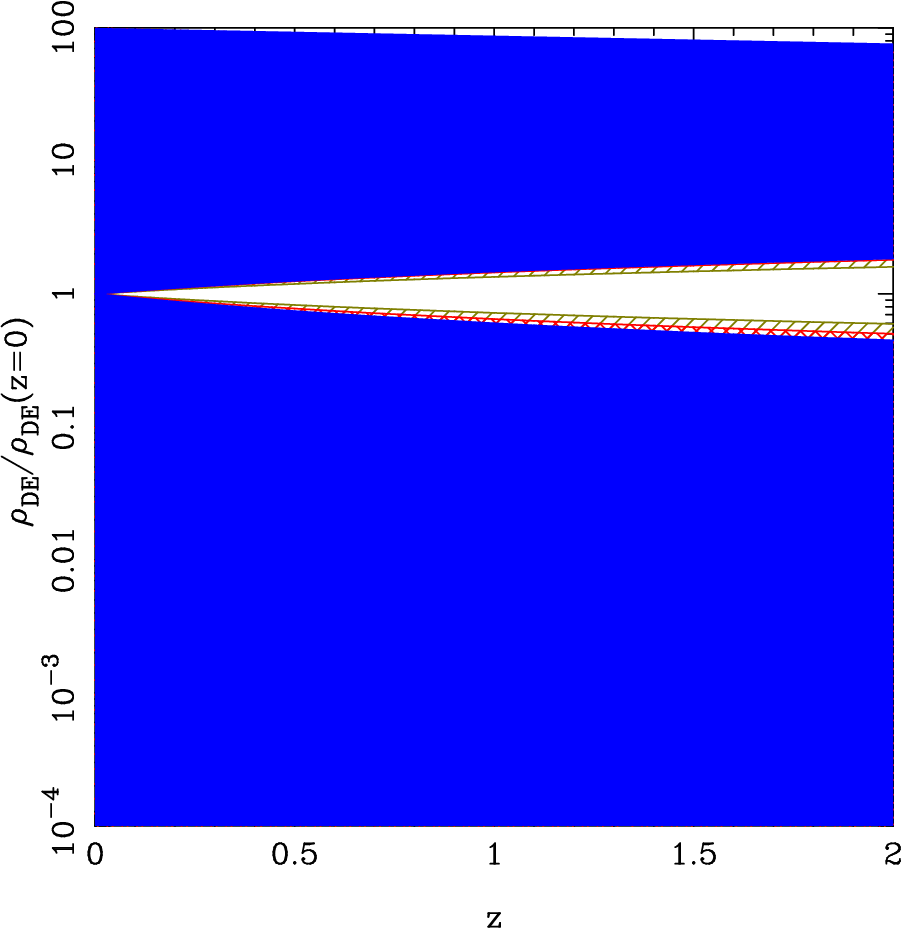}
\\
\includegraphics[width=1.9in]{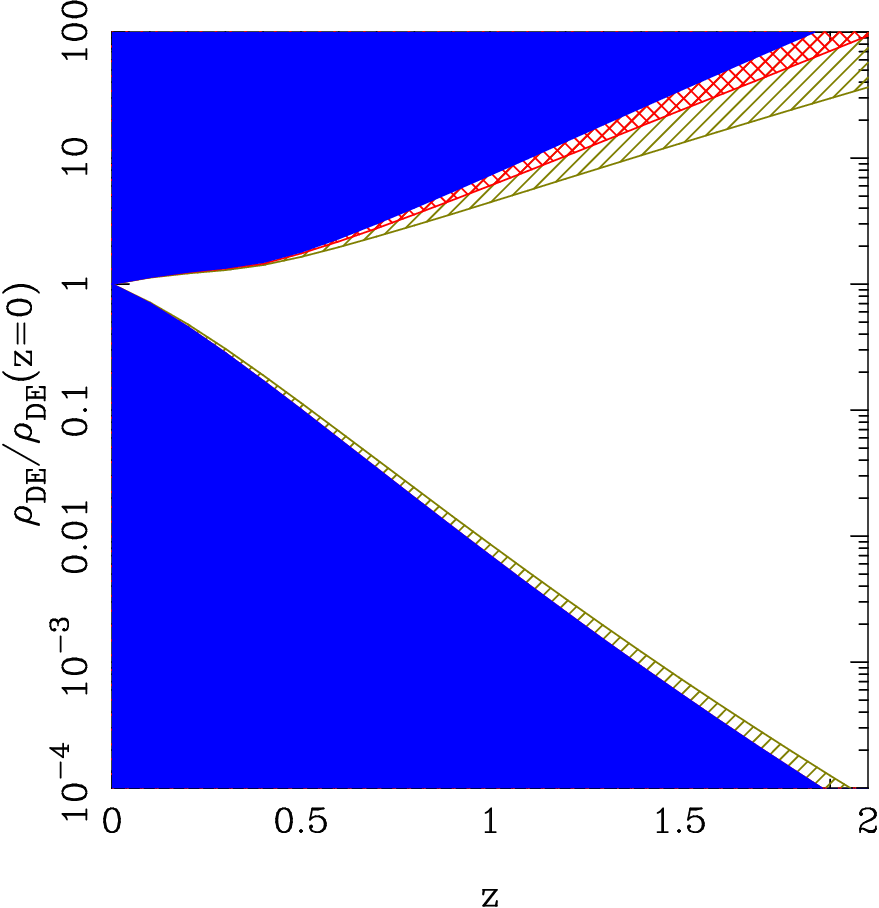} 
&
\includegraphics[width=1.9in]{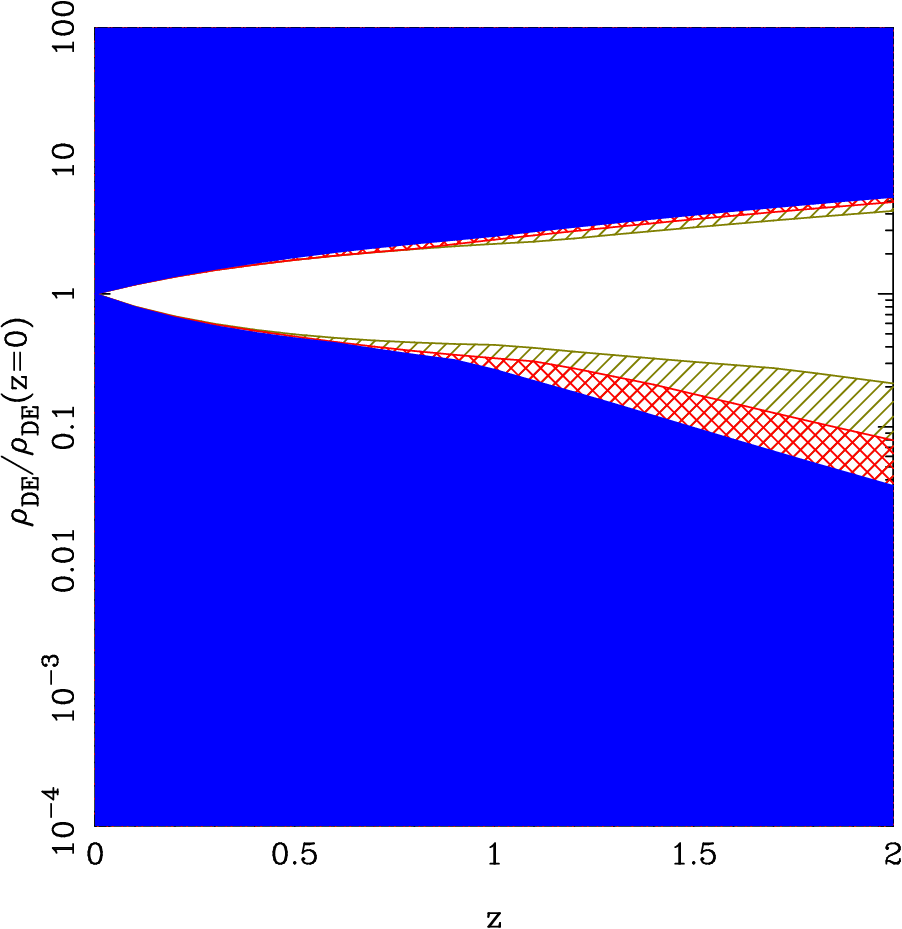}
&
\includegraphics[width=1.9in]{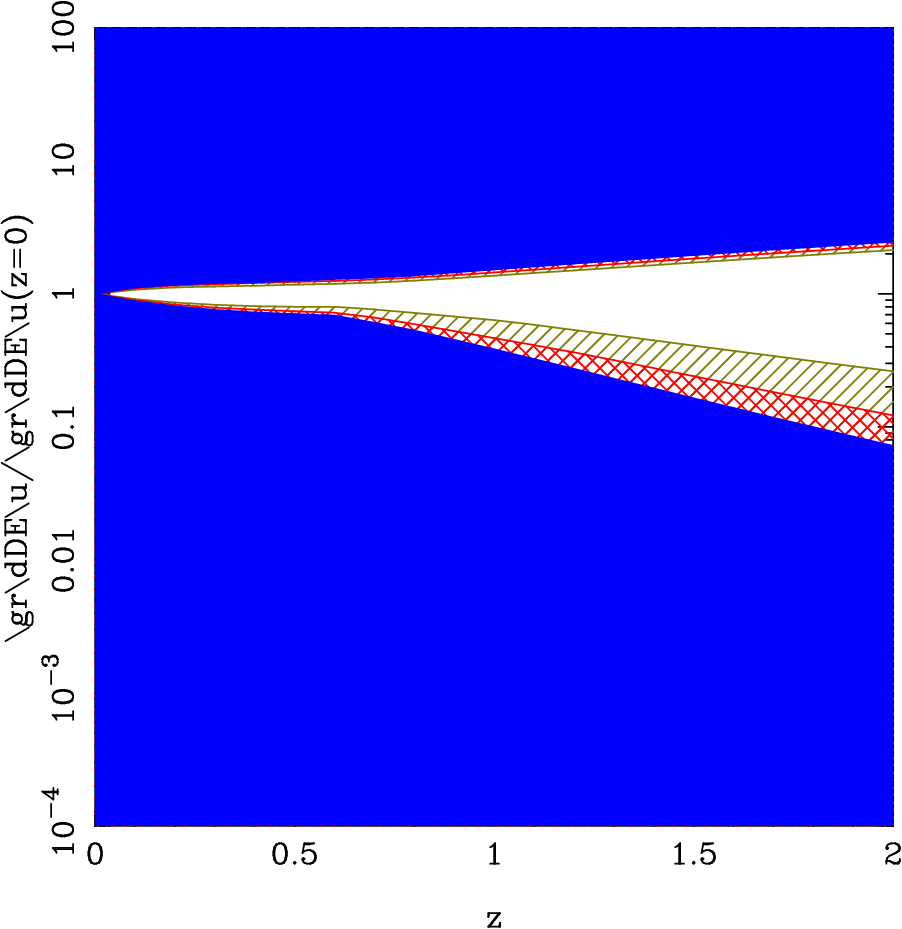}
\\
\includegraphics[width=1.9in]{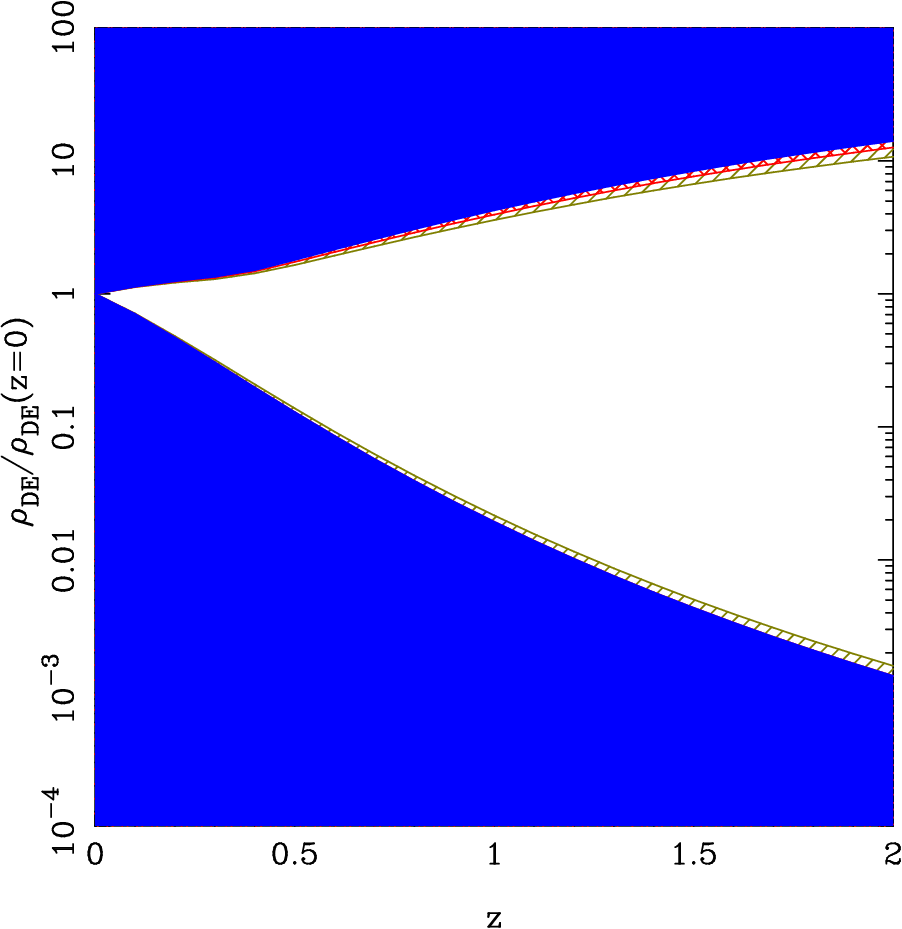} 
&
\includegraphics[width=1.9in]{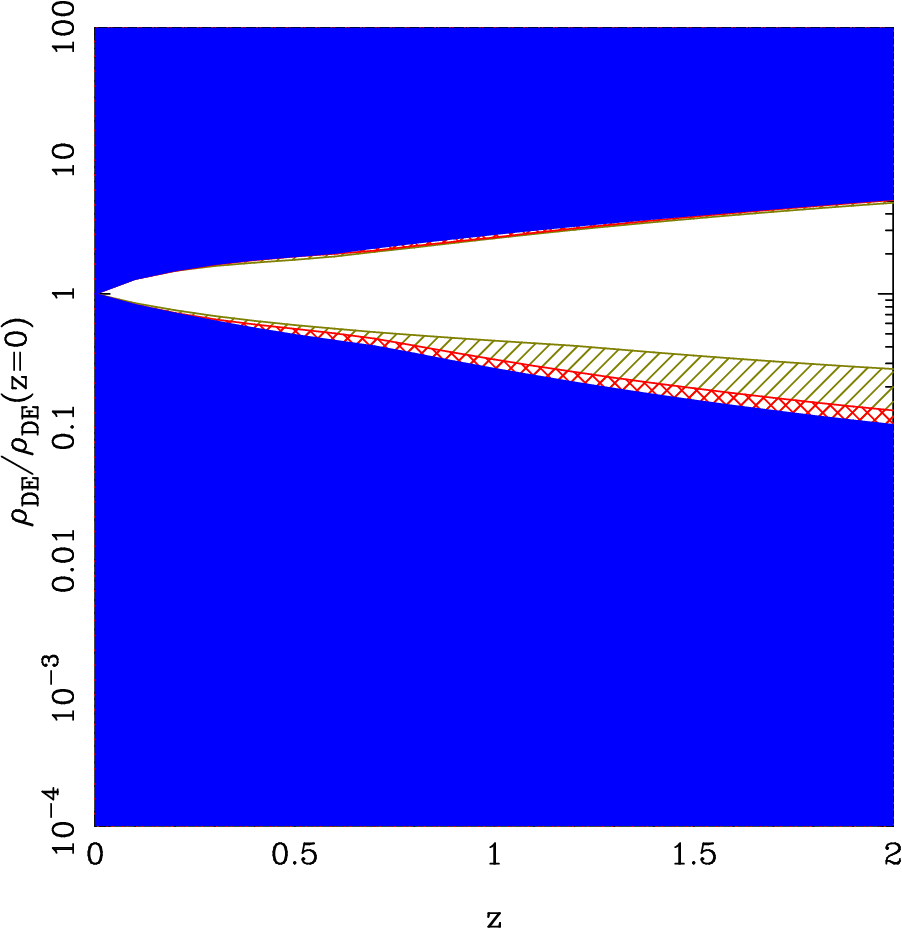}
&
\includegraphics[width=1.9in]{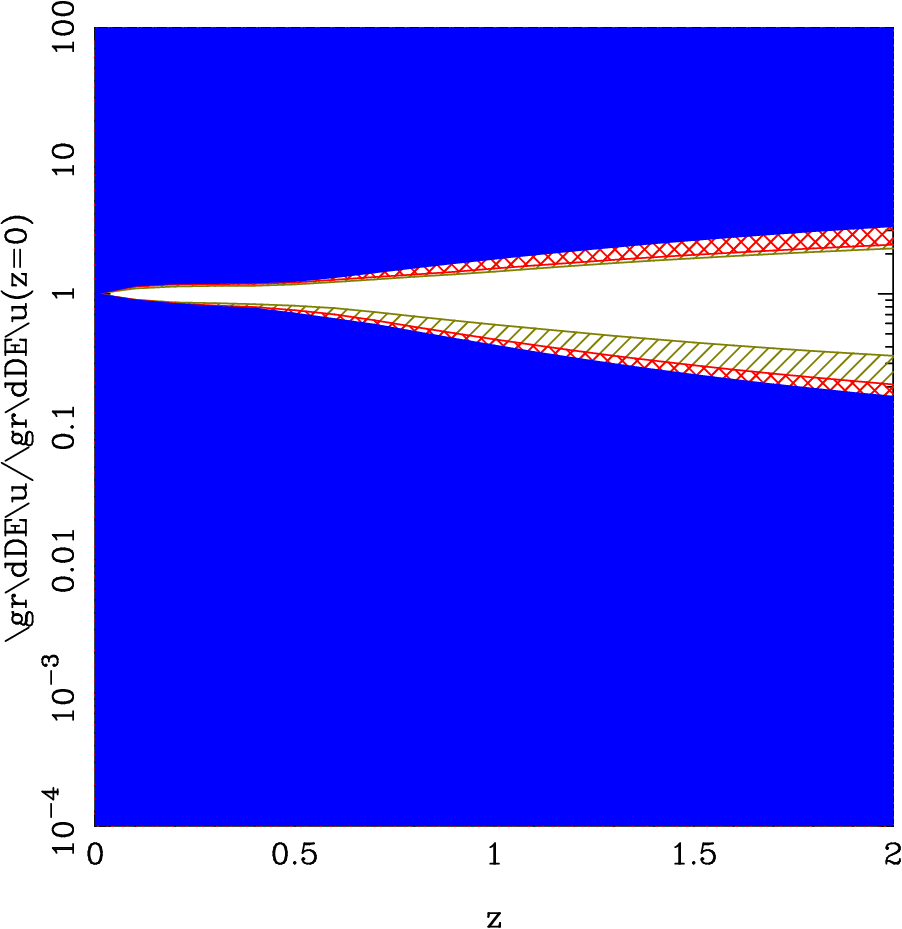}
\end{tabular}
\end{center}
\caption{This figure shows allowed range in variation of dark energy density
  as a function of redshift. The top row is for homogeneous dark energy model,
  the   second row is for perturbed dark energy, the third row is for varying
  $w$   with $p=1$, and, the last row is for varying $w$ with $p=2$.  
 The white region is the allowed range at 63\% confidence level, the hatched
 region is the one disallowed range of dark energy density at 95\% confidence
 level and the solid (blue) region is the one ruled out at 99\% confidence
 level. In all the rows, the left most plot shows range allowed by SNLS data,
  the middle one with WMAP5 data and the right one shows the range allowed by
  combined data.} 
\end{figure*}
%%%%%%%%%%%%%%%%%%%%%%%%%%%%%%%%%%%%%%%%%%%%%%%%%%%%%%%%%%%%%%%%%%%%%%%
%%%%%%%%%%%%%%%%%%%%%%%%%%%%%%%%%%%%%%%%%%%%%%%%%%%%%%%%%%%%%%%%%%%%%
\begin{table*}
\label{tab:range2}
\caption{This table lists the range of parameters allowed within 95\%
  confidence limit from SNLS,  WMAP1, WMAP3 and SNLS+WMAP3.} 
\begin{center}
\begin{tabular}{||l|c|c|c|c|c|c||}
\hline
\hline
Parameter   & $\Lambda$CDM  &     w=const. & w=const.                   &p=1         & p=2 &\\
      &         &              & with perturbations         &            &
&\\            
\hline
\hline
            &  &$-1.92$ --- $-0.74$  &$-1.92$---$-0.74$ &$-1.89$---$-0.61$ &$-1.9$---$-0.59$ & SNLS \\
$w$         &    &$-1.39$ --- $-0.58$ & $-1.63$ --- $-0.66$&$-1.64$ --- $-0.42$
&$-1.93$ --- $-0.43$ & WMAP1 \\
%         &    &$-1.24$ --- $-0.66$ & $-1.25$ --- $-0.61$&$-1.6$ --- $-0.4$ &$-1.6$ --- $-0.43$ & WMAP3 \\    
         &    &$-1.25$ --- $-0.7$ & $-1.25$ --- $-0.64$&$-1.62$ --- $-0.44$ &$-1.62$ --- $-0.43$ & WMAP5 \\
            & & $-1.47$ --- $-0.83$&$-1.57$ --- $-0.88$  &$-1.46$ --- $-0.81$ & $-1.74$ --- $-0.77$& SNLS+WMAP1 \\    
%            & & $-1.48$ --- $-0.59$&$-1.87$ --- $-0.57$  &$$ --- $$ & $$ ---
%            $$& SNLS+WMAP3 \\
            & & $-1.1$---$-0.9$&$-1.1$ --- $-0.9$  & $-1.3$ --- $-0.8$ &$-1.42$ --- $-0.66$& SNLS+WMAP5 \\
\hline
\hline
             &&  & &$-4.82$---$3.3$ &$-4.79$---$4.23$ &SNLS \\
$w'(z=0)$     &       &  & &$-3.09$ --- $1.32$ &$-2.5$ --- $4.87$ &WMAP1\\
%               &       &  & &$-1.7$ --- $1.2$ &$-3.1$ --- $2.8$ &WMAP3\\
               &       &  & &$-1.8$ --- $1.2$ &$-3.3$ --- $2.7$ &WMAP5\\
             &&  & &$-0.99$ --- $1.04$ & $-2.22$ --- $4.79$& SNLS+WMAP1 \\    
%             &&  & &  ---  &  --- & SNLS+WMAP3 \\ 
             &&  & & $-1.25$ --- $0.91$ & $-2.6$ --- $2.7$ & SNLS+WMAP5 \\ 
\hline
\hline
             &$0.22$---$0.31$ &$0.11$---$0.47$  &$0.11$---$0.47$ &$0.11$---$0.48$ &$0.11$---$0.48$& SNLS \\
$\Omega_{NR}$ &$0.20$ --- $0.45$ &$0.16$ --- $0.43$   &$0.20$ ---
$0.47$&$0.17$ --- $0.45$ &$0.18$ --- $0.44$& WMAP1 \\
&$0.19$ --- $0.33$ &$0.19$ --- $0.38$   &$0.19$ --- $0.39$&$0.19$ --- $0.39$
&$0.19$ --- $0.38$& WMAP3 \\
&$0.21$ --- $0.34$ &$0.2$ --- $0.38$   &$0.2$ --- $0.38$&$0.2$ --- $0.38$ &$0.20$ --- $0.38$& WMAP5 \\
             &$0.22$---$0.31$ & $0.15$---$0.36$ &$0.18$ --- $0.41$ &$0.16$---$0.38$ &$0.19$---$0.39$ & SNLS+WMAP1 \\    
% &$0.19$---$0.42$ & $0.17$---$0.43$ &$0.15$ --- $0.43$ & ---  & --- & SNLS+WMAP3 \\
& $0.22$ --- $0.3$ & $0.22$ --- $0.3$ & $0.22$ --- $0.3$ & $0.22$ --- $0.33$  & $0.21$ --- $0.32$ & SNLS+WMAP5 \\
\hline
\hline
$h$         &$0.61$ --- $0.79$  &$0.61$ --- $0.78$  &$0.6$ ---
$0.79$& $0.6$ --- $0.78$  & $0.61$ --- $0.78$ & WMAP1\\
%&$0.66$ --- $0.78$  &$0.60$ --- $0.79$  &$0.6$ ---
%$0.79$& $0.6$ --- $0.78$  & $0.61$ --- $0.79$ & WMAP3\\
&$0.66$ --- $0.77$  &$0.60$ --- $0.78$  &$0.60$ ---
$0.78$& $0.61$ --- $0.79$  & $0.61$ --- $0.78$ & WMAP5\\
            & $0.69$---$0.77$ & $0.68$---$0.78$ & $0.68$---$0.79$ &
$0.67$---$0.77$&$0.65$---$0.78$& SNLS+WMAP1 \\ 
%    & $0.61$---$0.79$ & $0.61$---$0.79$ & $0.61$---$0.79$ & --- &---& SNLS+WMAP3 \\
    & $0.68$---$0.74$ & $0.68$---$0.74$ & $0.68$---$0.74$ & $0.66$ --- $0.76$
& $0.65$ --- $0.76$ & SNLS+WMAP5 \\
\hline
\hline

$\Omega_B h^2$& $0.02$ --- $0.027$ & $0.021$ --- $0.028$ &$0.02$ ---
$0.027$&$0.02$ --- $0.027$ &$0.02$ --- $0.027$& WMAP1\\
%& $0.02$ --- $0.024$ & $0.021$ --- $0.024$ &$0.02$ ---
%$0.024$&$0.02$ --- $0.024$ &$0.02$ --- $0.024$& WMAP3\\
& $0.021$ --- $0.024$ & $0.021$ --- $0.023$ &$0.02$ ---
$0.024$&$0.021$ --- $0.0235$ &$0.022$ --- $0.024$& WMAP5\\
            & $0.02$ --- $0.027$ & $0.021$---$0.027$ &$0.021$---$0.027$
&$0.021$---$0.027$ &$0.021$---$0.028$& SNLS+WMAP1 \\     
%&$0.93$ --- $1.03$ & $0.9$---$1.04$ & $0.89$---$1.04$& ---  & ---& SNLS+WMAP3 \\ 
&$0.021$ --- $0.023$ & $0.021$---$0.024$ & $0.021$---$0.024$ & $0.021$ ---
$0.023$  &  $0.021$ --- $0.024$ & SNLS+WMAP5 \\ 
\hline
\hline

$n$          &$0.93$ --- $1.08$  &$0.93$ --- $1.1$   &$0.93$ ---
$1.09$&$0.93$ --- $1.1$   &$0.93$ --- $1.098$&WMAP1\\
%  &$0.93$ --- $0.99$  &$0.93$ --- $0.99$   &$0.93$ ---
%$1.0$&$0.93$ --- $0.99$   &$0.93$ --- $0.99$&WMAP3\\
  &$0.94$ --- $0.99$  &$0.93$ --- $0.99$   &$0.93$ ---
$0.99$&$0.93$ --- $0.99$   &$0.93$ --- $0.99$&WMAP5\\
            &$0.93$ --- $1.09$ & $0.94$---$1.08$ & $0.94$---$1.09$&$0.93$ --- $1.09$ &$0.93$ ---
$1.097$ & SNLS+WMAP1 \\     
% &$0.93$ --- $1.03$ & $0.9$---$1.04$ & $0.89$---$1.04$& $ $ --- $ $ &$ $ ---
%$ $ & SNLS+WMAP3 \\
 &$0.94$ --- $0.99$ & $0.94$---$0.99$ & $0.93$---$0.99$& $0.93 $ --- $0.99$ & $0.94$ ---
$0.99$ & SNLS+WMAP5 \\
\hline
\hline

$\tau$       &  $0.002$ --- $0.33$   &$0.011$ --- $0.39$  &$0.007$ ---
$0.35$ &$0.13$ --- $0.4$  &$0.016$ --- $0.39$ &WMAP1\\
%&  $0.027$ --- $0.15$   &$0.028$ --- $0.15$  &$0.027$ ---
%$0.15$ &$0.03$ --- $0.15$  &$0.027$ --- $0.15$ &WMAP3\\
&  $0.054$ --- $0.12$   &$0.05$ --- $0.12$  &$0.055$ ---
$0.12$ &$0.054$ --- $0.12$  &$0.054$ --- $0.12$ &WMAP5\\
            & $0.004$ --- $0.33$& $0.008$---$0.34$ &$0.045$---$0.37$ &$0.013$ --- $0.38$ &$0.017$ --- $0.396$ & SNLS+WMAP1 \\    
% & $0.034$ --- $0.22$& $0.008$---$0.22$ &$0.005$---$0.21$ &  --- 
%& ---  & SNLS+WMAP3 \\
 & $0.054$ --- $0.12$& $0.053$---$0.12$ &$0.055$---$0.13$ & $0.52$ --- $0.12$
& $0.05$ --- $0.125$ & SNLS+WMAP5 \\
\hline
\hline

\end{tabular}
\end{center}
\end{table*}
%%%%%%%%%%%%%%%%%%%%%%%%%%%%%%%%%%%%%%%%%%%%%%%%%%%%%%%%%%%%%%%%%%%%%%%%%%

We studied constraints on models of varying dark energy with the SNLS and
WMAP5 data and the results are summarised in table~2, which gives the ranges
of parameters allowed at $95\%$ confidence level.
The supernova data constrains $w_0$ but does not effectively
constrain $w_1$.
On the other hand, CMB data constrains an effective equation of state and hence
indirectly provides constraints on $w'(z=0)$. 
This is evident from lower panels of Figure~5, where we have plotted confidence
contours in $\Omega_{NR}-w$ plane.
The CMB contours are significantly narrower than those given by the Supernova
data. 
We do not find any significant changes in the likelihood contours for
other parameters such as $n$, $\Omega_{NR}h^2$, $\Omega_B h^2$ and $\tau$.
For instance the range of $\Omega_{NR}h^2$ is given by $0.21-0.33$ is valid
for all the models considered here.
This is illustrated in Table~2.

Given that CMB observations constrain only one number, we expect that the
constraint on models with varying $w(z)$ should constrain only $w_{eff}$ as
defined in Eqn.(\ref{eqn:weff}). 
We confirm this by plotting likelihood contours for models with varying $w(z)$
on the $\Omega_{NR}-w_{eff}$ plane.  
We also plot contours for constant $w$ models on the same plane for reference.
Figure~6 shows these contours for the models with $p=1$ and $p=2$.  
We find very strong coincidence in the contours for models with variable
$w(z)$ when plotted with $w_{eff}$ with the contours for models with constant
$w$ thereby confirming our conjecture. 
This also provides an easy approach to constraining models with variable
$w(z)$ without detailed calculations, all that one needs is to check whether
$w_{eff}$ is in the range allowed by CMB observations for $w$ in the constant
$w$ model.

We return to the issue of variation of dark energy density allowed by
observations. 
A pictorial representation of results is given in Figure~7, where we have
plotted $\rho_{DE}(z)/\rho_{DE}(z=0)$ as a function of redshift.  
Different panels show the evolution of this quantity as allowed by the
SNLS data set, WMAP5 observations of temperature and polarization 
anisotropies in the CMB and combined constraints from WMAP5+SNLS. 
These are plotted for constant $w$ (with and without dark energy
perturbations), and for variable $w$ with $p=1$ and $p=2$. 
Dark energy was assumed to be homogeneous in all cases except for the second
row that corresponds to constant $w$ models with perturbations in dark energy. 
We can conclude that: 
\begin{itemize}
\item
Supernova observations are a tight constraint for models with constant $w$,
but these are not as strong as CMB constraints.
\item
SNLS+WMAP5 data offers tighter constraints than either data set and the
cosmological constant is allowed with a high probability.  This follows from
the complementary nature of the two constraints as seen from the orientation
of the likelihood contours (e.g., see Figure~5).
\item
Supernova observations do not constrain evolution of dark energy density in
models with a variable $w$. 
Very large variation in dark energy density is allowed by these observations. 
\item
WMAP5 observations are, in contrast, a much tighter constraint and do not allow
significant variation in dark energy.  
Indeed, the variation in dark energy density allowed by WMAP5 observations for
models with variable $w$ is not significantly larger than that allowed for
constant $w$ models. 
\item
We demonstrate that the constraints on dark energy parameters for varying $w$
models are the same as the constraints on constant $w$ models if we consider
$w_{eff}$ for varying dark energy models. 
\item
SNLS+WMAP constraints are essentially dominated by the WMAP data and follow
the same pattern. 
SNLS observations add to the overall constraint by limiting the range of
values allowed for $w_0$.
\end{itemize}
We would like to add a note of caution that the analysis for varying $w$
models does not take perturbations in dark energy into account. 
However, these are more important for $w \gg -1$ or models with rapidly
varying $w$  \citep{perturb1,perturb2,perturb3,perturb4} and such models are
not allowed by observational constraints.  

\section{Discussion}

In this work we studied the SNLS data set and compared the
constraints obtained with constraints from WMAP five year data on temperature
anisotropies in the CMB.   
We find that the parameter values favoured by the two data sets have
significant overlap and the two sets can be combined to put tight constraints
on models of dark energy.   
In an earlier work we had noted that the Gold+Silver data set does not agree
with WMAP observations in that these favour distinct parts of the parameter
space \citep{2005PhRvD..72j3503J}.   
Constraints from WMAP and structure formation favour similar models, but
ones distinct from those favoured by Gold+Silver supernova observations.
This indicates some degree of inconsistency between the supernova and other
observations and it led us to suggest that the Gold+Silver data set may be
affected by as yet unknown systematic errors \citep{2005PhRvD..72j3503J}.
One reason for doubting the supernova data is the heterogeneity of sources
from which the particular data set was collected \citep{nova_data3}.  
SNLS \citep{snls} is a homogeneous data set and should not suffer from such
problems and indeed we find that there is no inconsistency between SNLS and
WMAP observations.

This highlights the usefulness of CMB observations for constraining models of
dark energy \citep{white,jbp}.  
We believe that CMB observations should be used for testing any model of dark
energy as supernova observations do not constrain models with varying $w$
effectively \footnote{The main constraint on varying $w$ models is from the
  CMB data}. 
Thus one should use CMB observations as well and not rely only on supernova
observations for constraining such models. 

We would like to add a note of caution against combining the SNLS data
with other data sets of high redshift supernovae in light of the very
different nature of these data sets.  
Indeed, one should use homogeneous data sets like the SNLS in isolation  to
avoid the problems mentioned above.

In terms of models, we find that the cosmological constant is favoured by 
individual observations (SNLS and WMAP) as well as in the combined data set
with very high probability.
Table~2 gives allowed values of all cosmological parameters at $95\%$
confidence level by SNLS, WMAP as well as the combined data set.  
For the cases where a similar analysis has been done by others, our results
are consistent with other findings
\citep{2003Sci...299.1532B,wmap_params,constraints_5,constraints_7,constraints_8,2004ApJ...606..702T,constraints_2,constraints_13,constraints_14,constraints_3,constraints_10,constraints_9,constraints_10a,constraints_11,constraints_12a,constraints_12,peri_snls,2005PhRvD..71j3515S,constraints-12,feng1,feng2,feng3,
2007PhRvD..75h3506A,2006PhRvD..74h3519C,2006PhRvD..74b3532C,2007A&A...467..421D,2008arXiv0803.1311E,mota_constr}.   

We have discussed the origin of the constraint on dark energy models from CMB
observations at length.  
We may conclude from the analysis presented here that:
\begin{itemize}
\item
Location of acoustic peaks in the angular power spectrum of CMB anisotropies
is the main source of constraints. 
\item
CMB observations only constrain $w_{eff}$, an effective value of the equation
of state parameter defined in Eqn.(\ref{eqn:weff}). 
This can be used to translate constraints on models with constant $w$ to
models where dark energy properties vary with time. 
\item
We have discussed models with $\Omega_0 = 1$ in this paper.  
In case this constraint is relaxed then the well known degeneracy between $w$
and $\Omega_0$ loosens the constraints.
However, it is well known that the SN and CMB data are complementary and can
be combined to provide fairly tight constraints even in this case
\citep{lss_de,2001PhRvD..64l3527H}. 
We do not expect variations in curvature to modify our conclusions about
$w_{eff}$ being the only dark energy related quantity constrained by CMB
observations. 
\item
Integrate Sachs-Wolfe (ISW) effect due to perturbations in dark energy can, in
principle, lead to variations in the CMBR angular power slectrum at small
$l$.  
Our analysis of models with variable $w$ without perturbations does not take
this into account. 
Various analyses have shown that ISW does not contribute significantly to
constraining cosmological parameters from CMB data (see for example,
\citep{2009JCAP...09..003X}).  The main reason for this is that ISW affects
power at small $l$, and due to large cosmic variance these modes do not
contribute much to the overall likelihood. 
\end{itemize}

\section*{Acknowledgements}

Numerical work for this study was carried out at cluster computing
facility in the Harish-Chandra Research Institute
(http://cluster.hri.res.in).  
This research has made use of NASA's Astrophysics Data System.

\label{lastpage}

\end{document}